\documentclass[]{aa} 

\usepackage{graphicx}
\usepackage{txfonts}
\begin{document}

   \title{Opacity broadening and interpretation of suprathermal CO linewidths: Macroscopic Turbulence and Tangled Molecular Clouds}
  \titlerunning{Opacity broadening and interpretation of suprathermal CO linewidths}

   \author{A. Hacar
          \inst{1}
          \and
          J. Alves\inst{1}
          \and
          A. Burkert\inst{2,3}
          \and
          P. Goldsmith\inst{4}
          }

   \institute{Department of Astrophysics, University of Vienna,
              T\"urkenschanzstrasse 17, A-1180 Vienna, Austria\\
              \email{alvaro.hacar@univie.ac.at} 
              \and
              University Observatory Munich (USM), Scheinerstrasse 1, 81679 Munich, Germany \and
              Max-Planck-Fellow, Max-Planck-Institut f\"ur extraterrestrische Physik (MPE), Giessenbachstr. 1, 85748 Garching, Germany
              \and
              Jet Propulsion Laboratory, California Institute of Technology, Pasadena, CA 91109, USA
             }

   \date{XXX-XXX}

\abstract{Since their first detection in the ISM, (sub-)millimeter line observations of different CO isotopic variants have routinely been employed to characterize the kinematic properties of the gas in molecular clouds. Many of these lines exhibit broad linewidths that greatly exceed the thermal broadening expected for the low temperatures found within these objects. These observed suprathermal CO linewidths are assumed to be originated from the presence of unresolved supersonic motions inside clouds.}
{The lowest rotational J transitions of some of the most abundant CO isotopologues, $^{12}$CO and $^{13}$CO, are found to present large optical depths. In addition to well-known line saturation effects, these large opacities present a non-negligible contribution to their observed linewidths. 
Typically overlooked in the literature, in this paper we aim to quantify the impact of these opacity broadening effects on the current interpretation of the CO suprathermal line profiles.}
{Combining large-scale observations and LTE modeling of the ground J=1-0 transitions of the main $^{12}$CO, $^{13}$CO, C$^{18}$O isotopologues, we have investigated the correlation of the observed linewidths as a function of the line opacity in different regions of the Taurus molecular cloud.
}
{Without any additional contributions to the gas velocity field, a large fraction of the apparently supersonic (${\cal M}\sim$~2-3) linewidths measured in both $^{12}$CO and $^{13}$CO (J=1-0) lines can be explained by the saturation of their corresponding sonic-like, optically-thin C$^{18}$O counterparts assuming standard isotopic fractionation. Combined with the presence of multiple components detected in some of our C$^{18}$O spectra, these opacity effects seem to be also responsible of most of the highly supersonic linewidths (${\cal M}>$~8-10) detected in some of the broadest $^{12}$CO and $^{13}$CO spectra in Taurus. 
}
{Our results demonstrate that most of the suprathermal $^{12}$CO and $^{13}$CO linewidths reported in nearby clouds like Taurus could be primarily created by a combination of opacity broadening effects and multiple gas velocity components blended in these saturated emission lines. Once corrected by their corresponding optical depth, each of these gas components present transonic intrinsic linewidths consistently traced by the three $^{12}$CO, $^{13}$CO, and C$^{18}$O isotopologues with differences within a factor of 2. 
Highly correlated and velocity-coherent at large scales, the largest and highly supersonic velocity differences inside clouds are generated by the relative motions between individual gas components. In contrast to the classical interpretation within the framework of microscopic  turbulence, this highly discretized structure of the molecular gas traced in CO suggest that the gas dynamics inside molecular clouds could be better described by the properties of a fully-resolved macroscopic turbulence.
}


   \maketitle
%

\section{Introduction}

The appropriate combination of its high abundance, large binding energy, and small dipolar momentum makes carbon monoxide (CO) an ideal tracer of the cold and dense molecular phase of the Interstellar Medium (ISM). Since its discovery, and thanks to a propitious rotational constant placing its lower-J transitions at (sub-)millimeter wavelengths, the CO and its isotopologues became the most observed molecules in radioastronomical studies. The first detection of the fundamental $^{12}$CO (J=1-0) line \citep{WIL70}, its main isotopic variants $^{13}$CO (1--0) and C$^{18}$O (1--0) lines \citep{PEN71} as well as their higher (J=2-1) counterparts \citep{PHI73}, were obtained after the opening of the millimeter window. These findings were rapidly followed by similar detections in nearby galaxies \citep{RIC75,SOL75,SAG91}. Nowadays, observations of the different rotational transitions of CO are routinely used to investigate the physical properties of molecular clouds \citep[e.g., ][]{UNG87,GOL08}, the internal structure of the Milky Way \citep[e.g., ][]{DAM01,SCO87,SOL87}, and the gas content of extragalactic sources \citep[e.g., ][]{SOL97,GRA05,LER08}. 

The analysis of the CO line emission profiles is regularly employed to characterize the dynamical state of gas inside molecular clouds. 
In particular, estimations on the line-of-sight velocity dispersion can be obtained from the observed CO linewidths $\Delta V$.
Since the first millimeter studies, most of the observed $^{12}$CO and $^{13}$CO linewidths are known to greatly
exceed the thermal broadening expected for the gas kinetic temperatures measured inside clouds ($\sim$~10~K). Instead of part of a global velocity structure \citep[either collapse, expansion, or rotation; e.g.][]{LIS74}, the origin of these {\it suprathermal} linewidths was explained by the contribution of random, non-thermal motions at small scales \citep{ZUC74}. The empirically determined velocity dispersion-size relationship found in different galactic clouds was early recognized to present strong similarities with the power-law dependency expected in a Kolmogorov-like cascade \citep{LAR81}. Consistently reported later in both local \citep{SOL87} and extragalactic sources \citep{BOL08}, the emergence of these suprathermal CO linewidths and Larson's scaling relationship are interpreted as signatures of the supersonic nature of the gas motions inside clouds and represent the observational foundation of the turbulent theory for these objects \citep[see][and references therein]{MAC04,ELM04}.

Compared to the broad, supersonic $^{12}$CO and $^{13}$CO line profiles, roughly equal contributions of both thermal and non-thermal motions are found in the observed C$^{18}$O linewidths in nearby clouds \citep{MYE83,VIL94,ONI96}. Originally thought to be restricted to cores at scales of 0.1~pc \citep{GOO98}, constant, (tran-)sonic-like C$^{18}$O linewidths have been found to be characteristic of filaments and fibers at scales of $\gtrsim$~0.5~pc \citep{HAC11,ARZ13,HAC13}. In some exceptional cases, similarly narrow CO linewidths seem to describe the internal gas kinematics inside entire filamentary clouds at multi-parsec scales \citep{HAC15}. 

The systematic differences between the observed linewidths of the main CO isotopologues, with $\Delta V$($^{12}$CO)~>~$\Delta V$($^{13}$CO)~$> \Delta V$(C$^{18}$O), are commonly assumed as an additional manifestation of the scale-dependency of the gas velocity dispersion in clouds \citep[e.g., Type 3 description on][]{GOO98}. This description oversimplifies the treatment of these distinct CO isotopologues as density selective gas tracers.
Although appealing, the above picture is complicated by the different optical depths affecting most of the $^{12}$CO and, to lesser extent, $^{13}$CO low-J rotational transitions. The saturation effects on the observed line intensities have been widely studied in the past \citep{GOL99,PIN08,PIN10}.
Similarly, although commonly underestimated in the literature, these opacity effects can also severely affect the observed linewidths in some of the optically thick CO emission lines \citep[e.g.][]{PHI79}. 
In this paper, we aim to explore and constrain the impact of this opacity broadening on the modern interpretation of the CO line profiles and its influence on the kinematic properties derived for the cold molecular gas forming the clouds. 

The structure of the paper is as follow. Section \ref{sec:CoG} presents the theoretical framework for line formation and the predictions for the growth of the observed CO linewidths as a function of the line opacity. 
In Sect.~\ref{sec:observations}, we compare these predictions with large-scale observations of the fundamental J=1--0 transition of three main $^{12}$CO, $^{13}$CO, C$^{18}$O isotopic variants of CO on the Taurus region. There, we compare the evolution of the apparently highly supersonic CO spectra a as function of the line opacity in the case of both single and multicomponent lines. Finally, and in Sect.~\ref{sec:discussion}, we explore the consequences of our findings on these CO suprathermal linewidths within the turbulent framework describing the internal dynamical properties of molecular clouds.


\section{Opacity broadening}

\subsection{Curve-of-growth}\label{sec:CoG}

\begin{figure}
   \centering
   \includegraphics[width=\columnwidth]{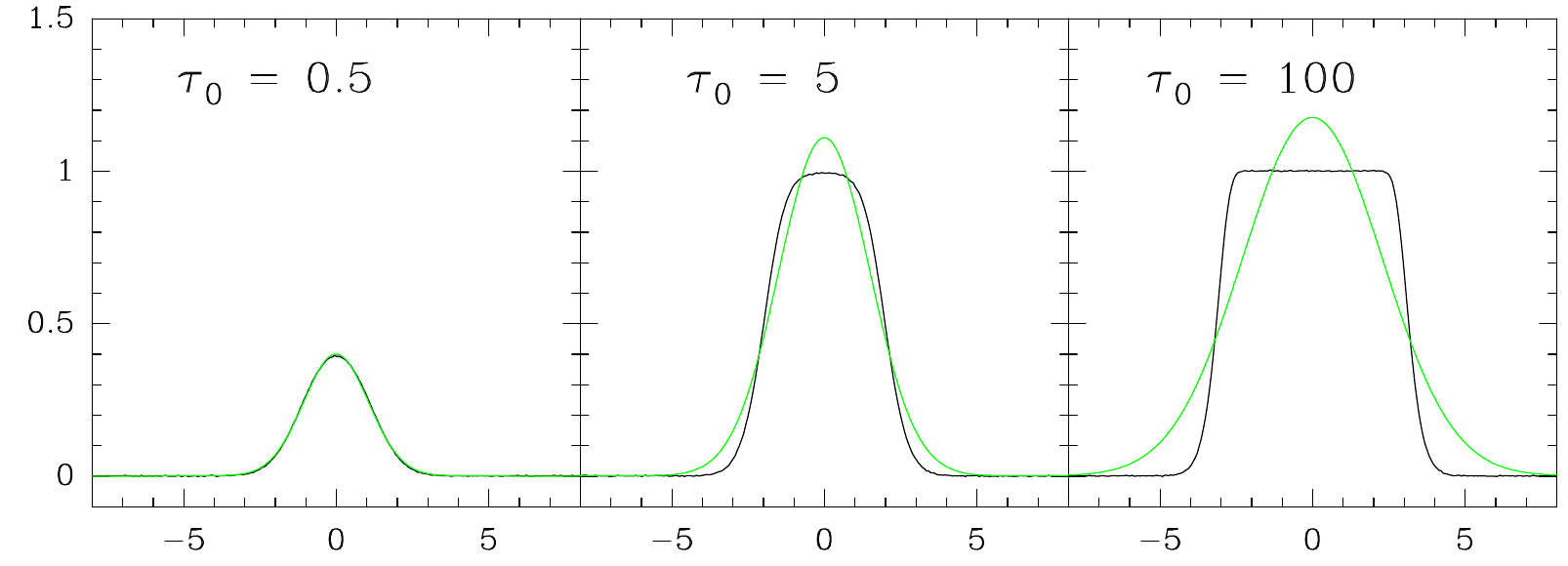}
   \caption{Line profiles (black lines) expected for an emission line described by an intrinsic gaussian-like velocity distribution and different central opacities: (Left) $\tau_0$=0.5; (Center) $\tau_0$=5; (Right) $\tau_0$=100. Line intensities and velocities are in arbitrary units. Note the saturation effects at opacities $\tau \ge 5$ resulting in top-flat line profiles deviating from a gaussian distribution (green lines).
   }
              \label{fig:thick_spectra}%
    \end{figure}

The spectral distribution of the emission of a molecular line is described by the radiative transfer equation \citep[e.g.][]{ROL96}:
\begin{equation}\label{eq:radiative_trans}
	T_{mb,\nu}=(J_\nu(T_{ex})-J_\nu(T_{bg}))\cdot(1-\exp(-\tau_\nu))
\end{equation}
With $J_\nu(T)=\left(\frac{h\nu/k}{exp(h\nu/kT)-1}  \right)$, T$_{ex}$ being the excitation temperature of the line, T$_{bg}$ the cosmic microwave background temperature (=2.7~K), and $\tau_{\nu}$ the line opacity at the line frequency $\nu$. 
For a gas with an intrinsic Maxwellian velocity distribution, the line opacity follows a gaussian distribution as: 
\begin{equation}\label{eq:tau_dist}
	\tau_{\nu}=\tau_0 \cdot exp(-(\nu-\nu_0)^2/2\sigma^2)
\end{equation}
Where $\tau_0$ and $\nu_0$ are the central opacity and line frequency, respectively, and $\sigma$ the intrinsic velocity dispersion of the gas. Knowing that for narrow lines  $\frac{\nu-\nu_0}{\nu_0} = \frac{V_{lsr}-V_{lsr,0}}{c}$, Eq.~\ref{eq:radiative_trans} and \ref{eq:tau_dist} can be used to predict the resulting line shape in velocity.  As illustrated in Fig.~\ref{fig:thick_spectra}, an evolution from gaussian-like to flat-topped and saturated spectra is expected in optically thick lines. 

The analysis of the observed line profiles can be used to infer the internal velocity field of the gas traced by a given molecular species. In the particular case of a optically-thin, gaussian-like molecular emission, the observed Full-Width-Half-Maximum (FWHM; $\Delta V$) is directly related to the gas internal velocity dispersion by $\sigma=\sqrt{8\ ln 2}\cdot \Delta V$. The intrinsic (Doppler) velocity dispersion of the gas ($\Delta V_{int}$ or $\sigma_{int}$) is produced by the joint contribution of both thermal ($\Delta V_{th}$ or $\sigma_{th}$) and non-thermal ($\Delta V_{NT}$ or $\sigma_{NT}$) motions along the line-of-sight added in quadrature:
\begin{equation}\label{eq:FWHM}
	\Delta V_{int}= \sqrt{ \Delta V_{th}^2 + \Delta V_{NT}^2 } \hskip0.2cm \longleftrightarrow \hskip0.2cm \sigma_{int} = \sqrt{ \sigma_{th}^2 + \sigma_{NT}^2 }
\end{equation}
where $\sigma_{th}=\sqrt{\frac{k\mathrm{T}_K}{m}}$ is the thermal velocity dispersion of a tracer with a molecular weight $m$ at a given gas kinetic temperature T$_K$. After the premise of $\Delta V_{int} = \Delta V$, Eq.~\ref{eq:FWHM} can be used to quantify the contribution of the non-thermal component of the the gas velocity field from the observed $\Delta V$ once T$_K$ is known. The magnitude of this non-thermal component is typically measured in units of the sound speed of H$_2$ at the same temperature as ${\cal M}=f(T_K)=\sigma_{NT}/c_s$ (e.g., $c_s=\sigma_{th}(H_2,10~K)=$~0.19~km~s$^{-1}$ or $\Delta V_{th}(H_2,10~K)=0.44$~km~s$^{-1}$, assuming $m=\mu_{H_2}$=2.33). As a key parameter describing the physical state of the gas inside clouds, the Mach number ${\cal M}$ is used to distinguish between the sonic (${\cal M}\le1$), transonic ($1<{\cal M}\le2$), and supersonic (${\cal M}> 2$) hydrodynamical regimes in non-magnetic, isothermal fluids\footnote{ In this paper, all $\sigma_{th}$, $\sigma_{NT}$, and $c_s$ are referred to as the 1D velocity dispersion of the gas along the line-of-sight. A factor of $\sqrt{3}$ is required to compare these values with the 3D velocity dispersions typically used in simulations (e.g., $c_{s,3D}=\sqrt{3}c_s=0.32$ km~s$^{-1}$). Their ratios compared to the sound speed (e.g., $\frac{\sigma_{NT,3D}}{c_{s,3D}}=\frac{\sigma_{NT}}{c_s}$) remain, however, unaltered.}.

   \begin{figure}[h]
   \centering
   \includegraphics[width=\columnwidth]{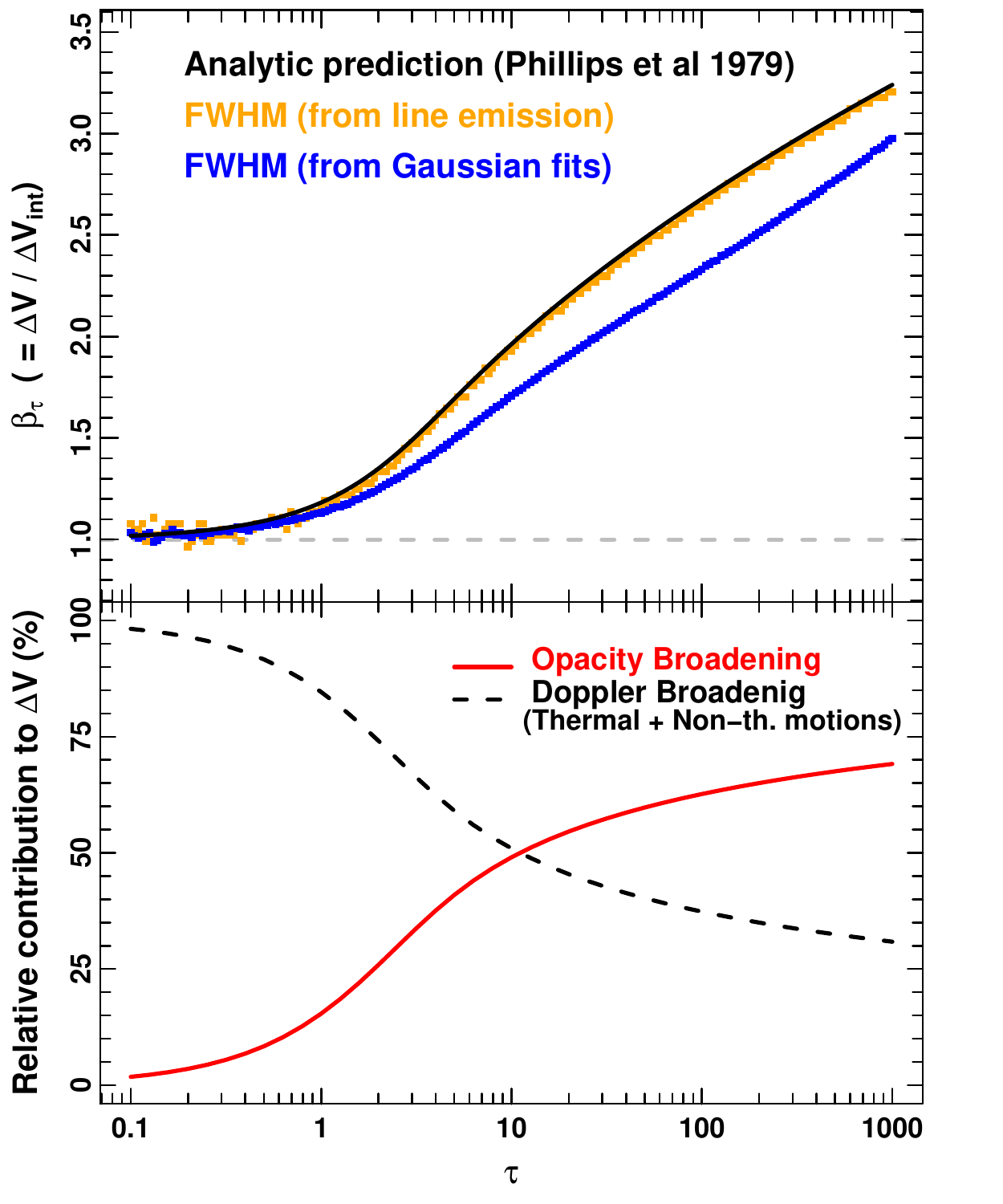}
   \caption{(Upper panel) Opacity broadening $\beta_\tau$ ($=\Delta V/ \Delta V_{int}$) as a function of the central opacity of the line $\tau_0$ (i.e., curve-of-growth): (Black) Analytic prediction \citep{PHI79};  (Orange) Values derived from the observed $\Delta V$ directly measured from the line emission; (Blue) Values derived $\Delta V$ from gaussian fits. (Lower panel) Contribution of both opacity (red- solid line) and intrinsic, Doppler broadening (thermal + non-thermal motions; black- dashed line) to the total observed linewidth for different line opacities.}
              \label{fig:broad}%
    \end{figure}

While suitable for optically thin lines, the above assumption of $\Delta V_{int} = \Delta V$ is no longer valid in the case of
top-flat and saturated spectra.
In analogy to the optical absorption lines, the so-called \emph{curve-of-growth} of molecular lines describes the line broadening produced by the increasing line opacity and can be defined analytically as \citep[e.g., ][]{PHI79}: 
\begin{equation}\label{eq:broad}
	\beta_\tau=\frac{\Delta V}{\Delta V_{int}} =\frac{1}{\sqrt{ln 2}}\left[ln \left( \frac{\tau_0}{ln\left(\frac{2}{exp(-\tau_0)+1} \right) } \right) \right]^{1/2}
\end{equation}
Where $\Delta V$ and $\Delta V_{int}$ are referred to the observed and the intrinsic FWHM, respectively. From the above expression, it is easy to show that the opacity broadening presents an asymptotic behavior approaching
\begin{equation}\label{eq:lim_broad}
	\beta_\infty = \frac{\Delta V}{\Delta V_{int}} \rightarrow \left( \frac{ln\ \tau_0}{ln\ 2} \right) ^{1/2}\  when \ \tau_0 \rightarrow  \infty \ .
\end{equation}
The effective opacity broadening described in Eq.~\ref{eq:broad} as a function of $\tau_0$ is displayed in Fig.~\ref{fig:broad} both in absolute (Upper panel) and relative terms (Lower panel).
As demonstrated by these figures, the opacity broadening $\beta_\tau$ represents a minor effect ($\lesssim$~15\%) for optically thin lines with $\tau\lesssim1$. However, its contribution increases monotonically at larger opacities. Indeed, the opacity broadening dominates the observed linewidths at opacities $\tau>$~10, reaching values of $\sim$~65-70\% for $\tau>$~100. \emph{In these last saturated lines, the observed half-power linewidth broadening do not reflect the
intrinsic dispersion of the gas, but the increased width due
to the relatively more intense line wings resulting from the high opacity of the
central portion of the line}.

\subsection{CO linewidths: Differential opacity broadening}\label{sec:COs}

The line opacity of a J=(j-i) transition ($\tau_\nu^{ji}$) is related to the total column density N of a molecule as:
\begin{equation}\label{eq:col_den}
	N = \frac{8\pi \nu_{ji}}{c^3}\frac{Q}{A_{ji}g_j}\frac{e^{E_j/kT_{ex}^{ji}} }{e^ {h\nu_{ji}/kT_{ex}^{ji}} -1} \cdot \int \tau_\nu^{ji} d\nu
\end{equation}
Where Q is the partition function, $A_{ji}$ the Einstein coefficient, $g_j$ the level multiplicity, and $\nu_{ji}$ the frequency of the transition considered.
Assuming similar and uniform LTE conditions along the line-of-sight, the opacities of a given J transition for the different CO lines can be estimated by their relative abundances as $\tau(A)\simeq X(A/B)\cdot \tau(B)$ \citep[e.g., ][]{MYE83}. Characteristic values of $\tau(C^{18}O)$ between 0.2 and 0.6 based on the study of the C$^{18}$O (1-0) transition are found in different studies of dark clouds \citep{MYE83,VIL94,VIL00,ONI96}. 
Assuming typical relative abundances for the main three CO isotopologues in the local ISM (i.e., X(C$^{18}$O):X($^{13}$CO):X($^{12}$CO) = 1:7.3:560 \citet{WIL94}) these opacities translate into characteristic line opacities ranging between $\tau(^{13}CO)\sim$1.5-4.4 and $\tau(^{12}CO)\sim$110-340 for the J=1-0 transitions of the most abundant CO isotopologue \citep[e.g.][]{PHI79,WON08}. Since $\Delta V = \beta_\tau \Delta V_{int}$, it is then clear that while the effects of opacity broadening can be neglected in the case of the optically thin C$^{18}$O (1-0) lines their contribution can severely affect the observed FWHM in both $^{13}$CO (1-0) and, in particular, the $^{12}$CO (1-0) optically thick line profiles. These differential opacity effects and their impact on the analysis of the gas kinematics are confirmed by recent simulations including radiative transfer calculations \citep{BEA13,COR14}. The derived parameters for the gas velocity field obtained from these last two lines could be strongly contaminated if their corresponding opacity broadening is not properly subtracted from the observed FWHM, $\Delta V$.

This differential line broadening selectively affects the distinct CO isotopologues. Some of these optical depth effects can be partially mitigated by observing higher-J CO transitions. Assuming similar LTE excitation conditions for all the $^{12}$CO lines and typical gas temperatures of 10~K, from Eq.~\ref{eq:col_den} it is easy to show that the $\tau(2-1)/\tau(1-0)\sim0.74$ and $\tau(3-2)/\tau(1-0)\sim0.24$. These similar excitation levels are however rarely satisfied under NLTE conditions due to the higher critical densities of these higher-J transitions.  Indeed, opacity broadening effects have been also reported in the study of  both (3-2) and (2-1) line transitions in $^{12}$CO and $^{13}$CO \citep{WIL99}. These line broadening effects are added to the well characterized optical depth correction needed to properly estimate the total column density of optically thick lines like $^{12}$CO (1-0) (i.e., $C=\frac{\tau}{1-exp(-\tau)}$; \citet{GOL99}). Combining all these effects, opacity appears as a major contaminant in most of the $^{12}$CO and $^{13}$CO observations in molecular clouds.

\subsection{Opacity dependent line blending}\label{sec:blending}

   \begin{figure}[h]
   \centering
   \includegraphics[width=\columnwidth]{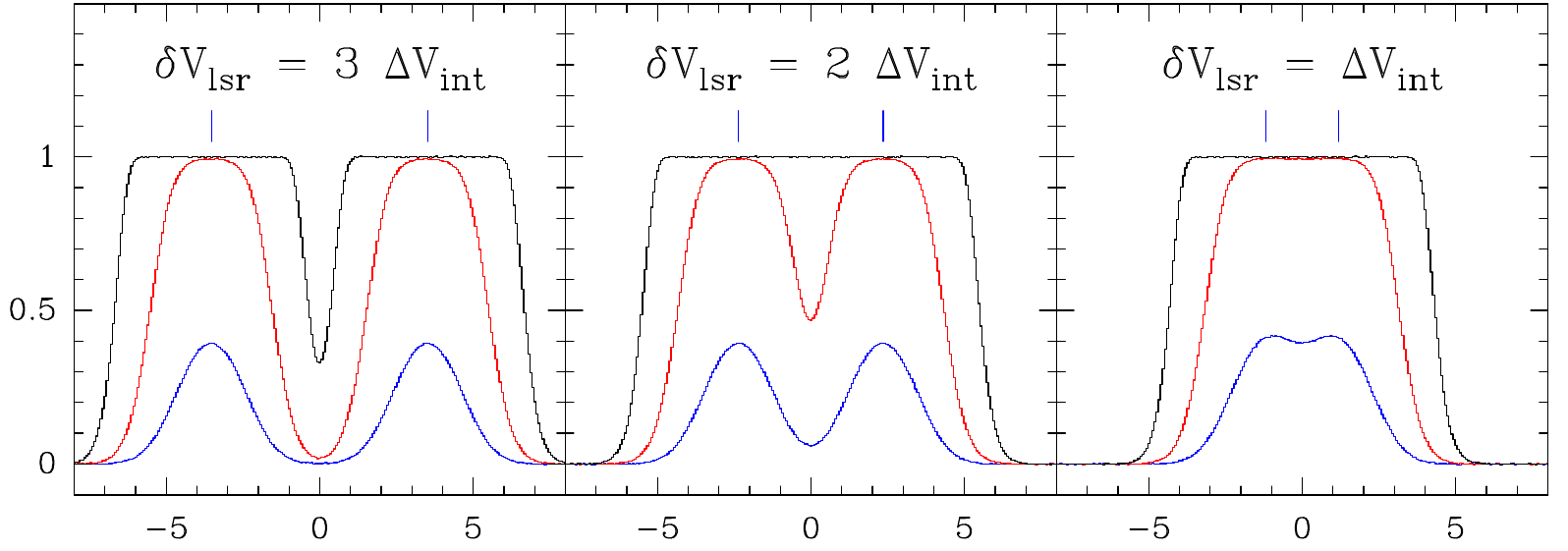}
   \caption{Line blending effects as a function of the velocity difference $\delta V_{lsr}$ between two adjacent components separated three (Left), two (Center), and one (Right) times their intrinsic linewidth $\Delta V_{int}$.
   Central line opacities of $\tau_0$=0.5 (blue), $\tau_0$=5 (red), and $\tau_0$=100 (black) are considered as representative of the three main CO isotopologues, that is, C$^{18}$O, $^{13}$CO, and $^{12}$CO, respectively. The central velocities of the two components forming each spectrum are indicated with blue lines.
   In the case of complex spectra with multiple components close in velocity ($\delta V_{lsr}\lesssim 2 \Delta V$), the individual lines can only be identified using optically thin tracers.
 }\label{fig:thick_blending}%
    \end{figure}

 As seen in Fig. 3, the opacity broadening also affects the ability to resolve the individual linewidth contributions to spectra with multiple components.
Two overlapping gaussian distributions can be independently resolved if their velocity difference $\delta V_{lsr}$ is larger than their corresponding $\Delta V$ \citep[see Apendix A in][]{HAC13}. In combination with Eq.~\ref{eq:broad}, this criterium yields a minimum velocity separation that is a function of the line opacity and the intrinsic velocity dispersion of the broadest component described as ${\cal R}=\delta V_{lsr}| _{min} = f(\tau,T_K,\sigma_{NT}) = \beta_\tau \Delta V_{int}$. 

It is easy to prove that the optical depth in thick tracers like $^{12}$CO or $^{13}$CO limits the characterization of the line substructure to those velocity components with larger separations in velocities, specially in warm regions. On the contrary, low opacity tracers like C$^{18}$O and C$^{17}$O are generally more appropriate for the study of complex lines with compact kinematic structures in velocity \citep{HAC11,HAC13} (see also Section~\ref{sec:supersonic}). As observational constrain, higher sensitivities are required to effectively detect such low abundant CO tracers.

\subsection{NLTE effects on the formation of CO lines}\label{sec:NLTE}

A core-to-edge temperature gradient is expected in molecular clouds produced by the incidence of the ISRF \citep[e.g.][]{YOU82}. As investigated by \citet{PIN10}, the resulting increase of T$_{ex}$ can effectively reduce the line opacities in the presence of large temperature gradients. Simultaneously, T$_{ex}$ is related to T$_K$ through collisional excitation. Thus, the increase of T$_K$ towards the cloud edges is counterbalanced by the likely reduction of the collisional rates at low densities leading to an sub-thermal T$_{ex}$ values.  The resulting line profile emerging from the cloud strongly depends on both its thermal structure and internal mass distribution. Indeed, the frequency dependence of the opacity makes different parts of the line become optically thick at different optical depths into the cloud and thus at different temperatures. As a result, large temperature and density variations along the line-of-sight could modify the optical depth and reduce or even suppress the formation of flat-topped profiles. The net impact of gradients on the observed line profiles remains unclear and might depend on the local properties of the cloud. First, systematic flat-topped $^{12}$CO and $^{13}$CO lines are identified in cold clouds with high contrasts in density like starless cores \citep{PAR04}. Conversely, smoother line profiles in optically thick lines like $^{12}$CO (1-0) are characteristic of regions with strong temperature contrasts \citep[e.g.][]{TAU91}. 

At low densities, photon trapping also plays an important role increasing T$_{ex}$  in optically thick transitions like the lower-J $^{12}$CO lines \citep{SCO74}. 
Due to the variations of the line opacity in velocity, the scape probability, and therefore the resulting emission arising from the cloud strongly depends on its geometrical shape and velocity structure. Overall, radiative trapping lowers the effective densities on which these optically thick lines are excited. In the case of the $^{12}$CO (1-0) transition, photon trapping is responsible for stimulating its line emission in gas at densities down to few 10$^2$~cm$^{-3}$, that is, an order of magnitude lower than in the absence of stimulated emission \citep{SHI15}. If these low densities are in regions of low extinction, however, the likely photodissociation of CO \citep{VDH88} complicates this picture. 

In addition to radiative transfer modeling, an adequate treatment of such effects requires an accurate description of both temperature and density structures inside clouds. Out of the scope of this work, none of the above NLTE effects are included in the analysis carried out in the following sections. Without loss of generality, the LTE treatment presented in this works aims to primary investigate the opacity effects on the observed CO linewidths. A comprehensive analysis of the CO emission and their opacity broadening under NLTE conditions will be presented in a future paper.
   
\section{CO observations in the Taurus molecular cloud}\label{sec:observations}

\subsection{Observed CO linewidths in the L1517 region}\label{sec:L1517}

According to Eq.~\ref{eq:broad}, and in the absence of additional kinematic effects, a line opacity broadening is expected to occur for lines with similar emission properties and increasing isotopic abundances. By comparing point-like observations of multiple CO transitions in different molecular clouds, \citet{PHI79} suggested that the linewidths differences observed between the distinct CO isotopologues follow the expected curve-of-growth of an emission line with increasing opacity. 
Continuing with this line of argument we have compared the observed line properties of the ground state J=(1-0) transition for the three main CO isotopologues in the L1517 region in Taurus \citep{LYN62}. 

   \begin{figure}[h]
   \centering
   \includegraphics[width=\columnwidth]{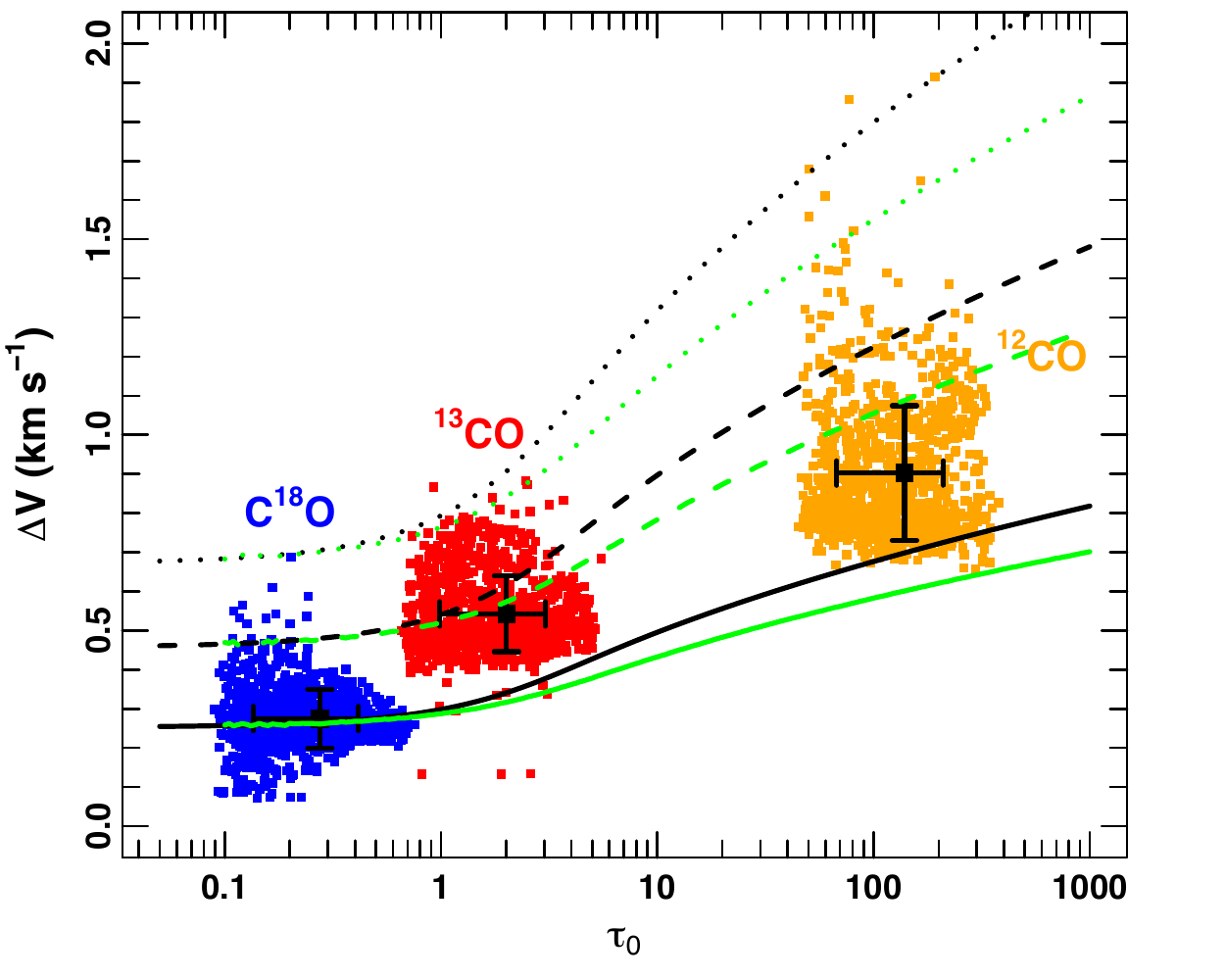}
   \caption{Observed C$^{18}$O (1-0) (blue), $^{13}$CO (red), and $^{12}$CO (1-0) (orange) linewidths ($\Delta V$) as a function of the central opacity ($\tau_0$) for all positions detected in L1517 with SNR(C$^{18}$O)~$\geq$~3 (> 940 points per molecule). The mean and standard deviation of these measurements of each isotopologue are indicated in the plot by black squares and bars.  
   The distinct curves correspond to the predicted opacity broadening of lines with an intrinsic Doppler broadening described by a thermal component at 10~K and a non-thermal velocity dispersion of 0.5~c$_s$ (solid line), 1.0~c$_s$ (long dashed line), and 1.5~c$_s$ (dotted line). In each case, this figure displays the comparison between the expected values for the opacity broadening derived from the analytic solution in Eq.~\ref{eq:broad} (black lines) with the measurements obtained from their corresponding gaussian fits to the spectra (green lines).}
              \label{fig:L1517_opa}%
    \end{figure}

The L1517 cloud was selected as it presents a simple gas velocity field as deduced from the global study of gas kinematics inside this region  \citep{HEY87,LAD91}. There, the global analysis of the C$^{18}$O (1-0) line emission at large scales shows the overwhelming presence of spectra with a unique and narrow gas velocity component \citep{HAC11}. We combine this large scale dataset with additional $^{13}$CO and $^{12}$CO (1-0) observations of the same region observed with the FCRAO14m radiotelescope during December 2003 and November 2005. The data reduction including flux calibration and main beam efficiency corrections reproduces the procedures described by \citet{HAC11} for the C$^{18}$O (1-0) lines. With similar coverage, the resulting spectral maps of the three CO isotopologues are convolved into a common Nyquist-sampled grid with a final beam-size of 60 arcsec.
To characterize their emission profile (i.e., T$_{mb}$, V$_{lsr}$, and $\Delta V$), the new $^{13}$CO and $^{12}$CO (1-0) spectra are parametrized using gaussian fits with the same number of components as their corresponding C$^{18}$O (1-0) counterpart (i.e., with one component in >~90\% of the observed spectra; see \citet{HAC11}). A total of 940 positions with SNR~$\ge$~3 in all the CO isotopologues are considered in the present study. The area sampled in this case is therefore restricted by the sensitivity of our C$^{18}$O maps. 

\begin{table*}[ht!]
\caption{Statistical properties of the observed CO spectra in L1517.}
\centering
\label{table:L1517}
\begin{tabular}{c|c|c|c||c|c}
\hline
\hline
	& C$^{18}$O  & $^{13}$CO  & $^{12}$CO & $^{13}$CO & $^{13}$CO  \\
\hline
Description & Standard  & Standard   & Standard   & Enhanced & Low T$_{ex}$  \\
J & (1-0) & (1-0) & (1-0) & (1-0) & (1-0) \\
X	&	1 &	7.3	 &	560	& 15 & 7.3  \\
T$_{ex}$	&	10 &	10	 &	10	& 10 & 8  \\
$\tau_0$  $^{(1)}$	&	0.28~$\pm$~0.14 &	2.05~$\pm$~1.03 &	158.7~$\pm$~80.4 & 4.20~$\pm$~2.11& 3.08~$\pm$~1.70   \\
$\langle \Delta V \rangle$~ (km~s$^{-1}$) &  0.28~$\pm$~0.07 & 0.54~$\pm$~0.09 &  0.90~$\pm$~0.17 &  0.54~$\pm$~0.09 &  0.55~$\pm$~0.09 \\
$\langle \Delta V_{int}\rangle$~ (km~s$^{-1}$)   $^{(2)}$ &  0.27~$\pm$~0.07 &  0.41~$\pm$~0.09  &  0.32~$\pm$~0.07 &  0.35~$\pm$~0.07 &  0.37~$\pm$~0.09  \\
$\langle \sigma_{NT}/c_s \rangle$ &  0.56~$\pm$~0.18 &  1.19~$\pm$~0.21 &  2.01~$\pm$~0.39	&  1.19~$\pm$~0.21 &  1.18~$\pm$~0.22 \\
$\langle \sigma_{NT,corr}/c_s \rangle$ &  0.53~$\pm$~0.18 &  0.87~$\pm$~0.21 &  0.67~$\pm$~0.16	&  0.73~$\pm$~0.18 &  0.79~$\pm$~0.20 \\
\end{tabular}
\tablefoot{ Uncertainties given correspond to 1$\sigma$ values.  $^{(1)}$ Line opacities obtained from Eq.~\ref{eq:tauC18O}. $^{(2)}$ Opacity corrected intrinsic linewidths estimated according to Eq.~\ref{eq:broad}. 

}
\end{table*}%
 
Based on the measured values for its main beam peak intensities T$_{mb}$, we directly estimate the opacity of C$^{18}$O (1-0) line at each position of our maps. By solving Eq.~\ref{eq:radiative_trans} assuming constant and uniform LTE excitation conditions at a typical temperature of T$_{ex}$=~T$_K=$~10~K the central opacity $\tau_0(C^{18}O)$ is calculated as 
\begin{equation}\label{eq:tauC18O}
	\tau_0(C^{18}O) = -log \left( 1- \frac{T_{mb}(C^{18}O)}{J(T_{ex})-J(T_{bg})} \right) \ .
\end{equation}
The corresponding opacities of the $^{12}$CO and $^{13}$CO isotopologues are then estimated at each point by scaling up the C$^{18}$O opacity according to their relative isotopic abundances, that is, $\tau(^{12}CO)=560\cdot\tau(C^{18}O)$ and  $\tau(^{13}CO)=7.3\cdot\tau(C^{18}O)$ (Sect.~\ref{sec:COs}). The statistical results for the line opacity of these three CO lines are summarized in Table~\ref{table:L1517}.

The combined analysis of the three independent datasets presented above allow us to systematically investigate of the opacity broadening of the three main CO isotopologues. Figure~\ref{fig:L1517_opa} displays the relation between the observed linewidths $\Delta V$ and the derived line opacities for all the positions with detected emission in all the C$^{18}$O, $^{13}$CO, and $^{12}$CO (1-0) lines in L1517. Overplotted to the observations, this figure shows the expected opacity broadening $\beta_\tau$ of three CO lines assumed to be described by an intrinsic Doppler velocity dispersion determined by a thermal component at 10~K and a non-thermal component of 0.5, 1.0, and 1.5 times the sound speed following Eq.~\ref{eq:FWHM}. As demonstrated by this comparison, the observed values for $\Delta V$ in the three main CO isotopologues show a systematic increase consistent with the expected curve-of-growth for an optically thick CO line with a non-thermal contribution $\sigma_{NT}$ between $\sim$~0.5 and 1.0~c$_s$. Table~\ref{table:L1517} also quantifies this result. While the observed $\Delta V$ differ a factor 2-3 between the two $^{13}$CO and $^{12}$CO (1-0) lines and their C$^{18}$O counterparts, these differences are reduced to $\lesssim~30$~\%
in the case of the opacity corrected $\Delta V_{int}$ values.

   \begin{figure}[h]
   \centering
   \includegraphics[width=\columnwidth]{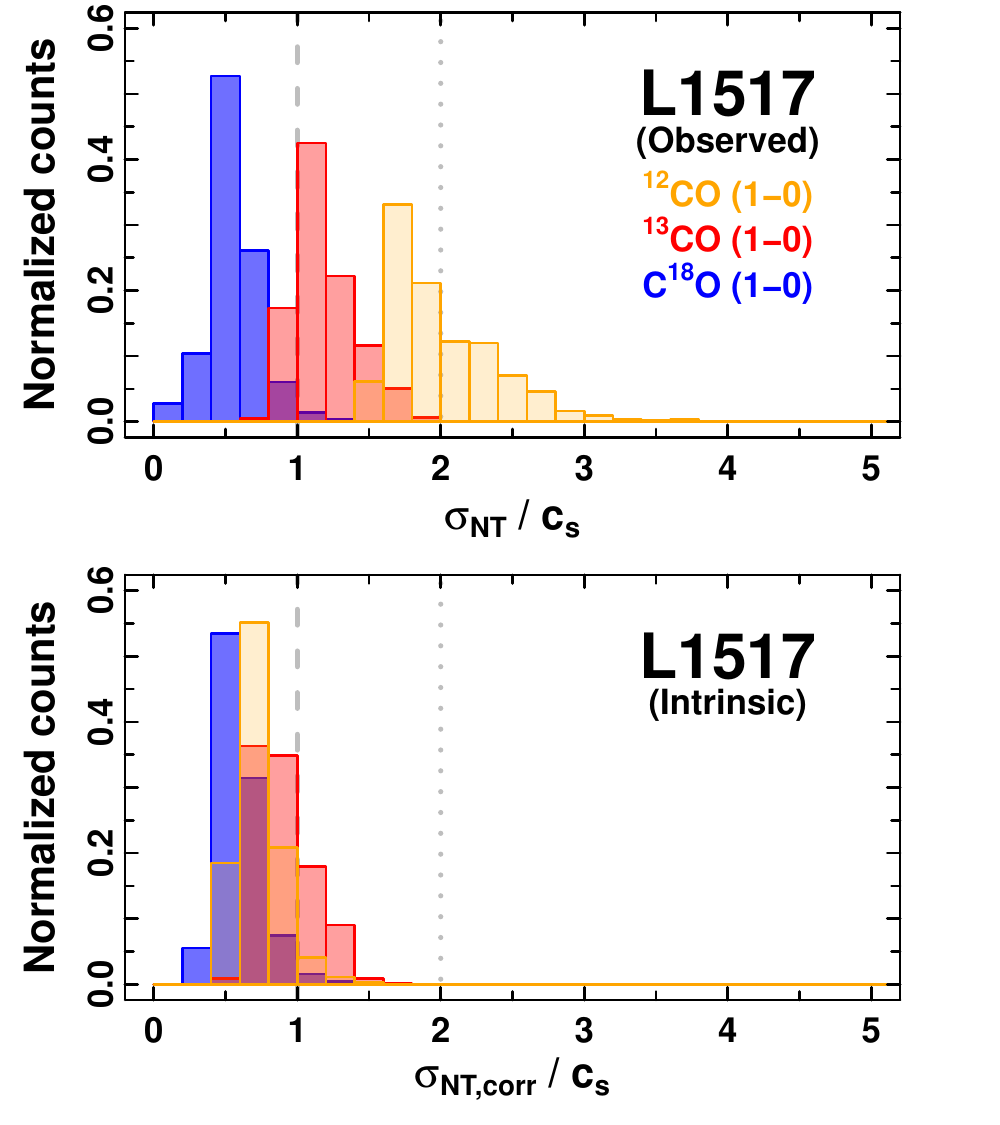}
   \caption{Histogram of the Non-thermal velocity dispersions (in units of the sound speed) measured in C$^{18}$O (1-0) (blue), $^{13}$CO (red), and $^{12}$CO (1-0) (orange) in L1517. (Upper panel) Non-thermal velocity dispersions directly derived from the observed linewidths ($\Delta V$). (Lower panel) Non-thermal velocity dispersions obtained from the intrinsic linewidths ($\Delta V_{int}$) after opacity corrections according to Eqs.~\ref{eq:broad} and \ref{eq:tauC18O}. We note that, although apparently supersonic, the $^{13}$CO and $^{12}$CO lines actually present sonic-line velocity dispersions. }
              \label{fig:L1517_histo}%
    \end{figure}

The inclusion of the opacity effects reported before have a strong impact on the interpretation of the parameters obtained from the observed CO linewidths. The histograms presented in Figure~\ref{fig:L1517_histo} (Upper panel) show the $\sigma_{NT}$ values derived from the observed $\Delta V$ for the three CO isotopologues (see also columns 1-3 in Table~\ref{table:L1517}). Taken as face values, these results would in principle suggest a tracer-dependent non-thermal velocity dispersion, varying from subsonic values of $ \langle \sigma_{NT}(C^{18}O)/c_s \rangle = 0.56\pm 0.18$, to transonic $\langle \sigma_{NT}(^{13}CO)/c_s \rangle = 1.19\pm 0.21$ and, ultimately, to supersonic $\langle \sigma_{NT}(^{12}CO)/c_s \rangle = 2.01\pm 0.39$ values. On the contrary, Fig.~\ref{fig:L1517_histo} (Lower panel) illustrates the distribution of the opacity corrected $\sigma_{NT,corr}$ using $\Delta V_{int}$ obtained from Eqs.~Eq.~\ref{eq:tauC18O} and \ref{eq:broad}. Once opacity effects are considered, the velocity dispersion of the gas derived from all of the CO isotopologues consistently yields systematically closer values with $\langle \sigma_{NT,corr}(C^{18}O)/c_s \rangle = 0.53\pm 0.18$, $\langle \sigma_{NT,corr}(^{13}CO)/c_s \rangle = 0.87\pm 0.21$ and $\langle \sigma_{NT,corr}(^{12}CO)/c_s \rangle = 0.67\pm 0.16$.  

While statistically significant differences can still be identified between the gas velocity field traced by the C$^{18}$O emission and the $^{13}$CO or $^{12}$CO lines, these differences are restricted to dynamical changes within the sonic regime. Excluded from our simplified analysis, the selective photodissociation of the C$^{18}$O molecules at different cloud depths \citep{BAL82} could explain the more dynamical velocity fields traced in $^{13}$CO and $^{12}$CO if small velocity gradients are present along the line-of-sight. Similar kinematic properties for these main CO isotopologues have been also reported in observations of quiescent regions like individual Bok globules \citep{DIC83} and the Musca cloud \citep{HAC15}.
In spite of these differences, the similar kinematics observed in the three main CO isotopologues demonstrates that these molecules are all sensitive to the same column density of molecular gas and that most of its gas kinematics is already mapped by the C$^{18}$O emission. Although other effects can not be ruled out, 
\emph{most of the differences in the observed linewidths between these tracers are consistent with pure opacity broadening effects}.

As secondary effect, systematically higher intrinsic velocity dispersions are derived from the $^{13}$CO emission in comparison with the $^{12}$CO lines in L1517. An underestimation of the $^{13}$CO line opacity seems to be the most plausible scenario explaining these differences \citep{PIN10}. Explored in the results shown in Table~\ref{table:L1517}, an increase of the derived line opacities for the $^{13}$CO (1-0) lines are expected if either the abundance of this molecule is enhanced by a factor of 2 compared to the local ISM values (column 4) and/or the line excitation temperatures drop down to 8 K (column 5). Both isotopic fractionation effects and sub-thermal excitation conditions have been commonly reported in the study in molecular clouds \citep{CHU83,LAN84,GOL08}. A NLTE analysis of the observations of multiple carbon monoxide isotopologues and rotational transitions is necessary to clarify this issue.

\subsection{Line blending and multiple velocity components in B213-L1495}\label{sec:supersonic}

   \begin{figure}[h]
   \centering
   \includegraphics[width=0.95\columnwidth]{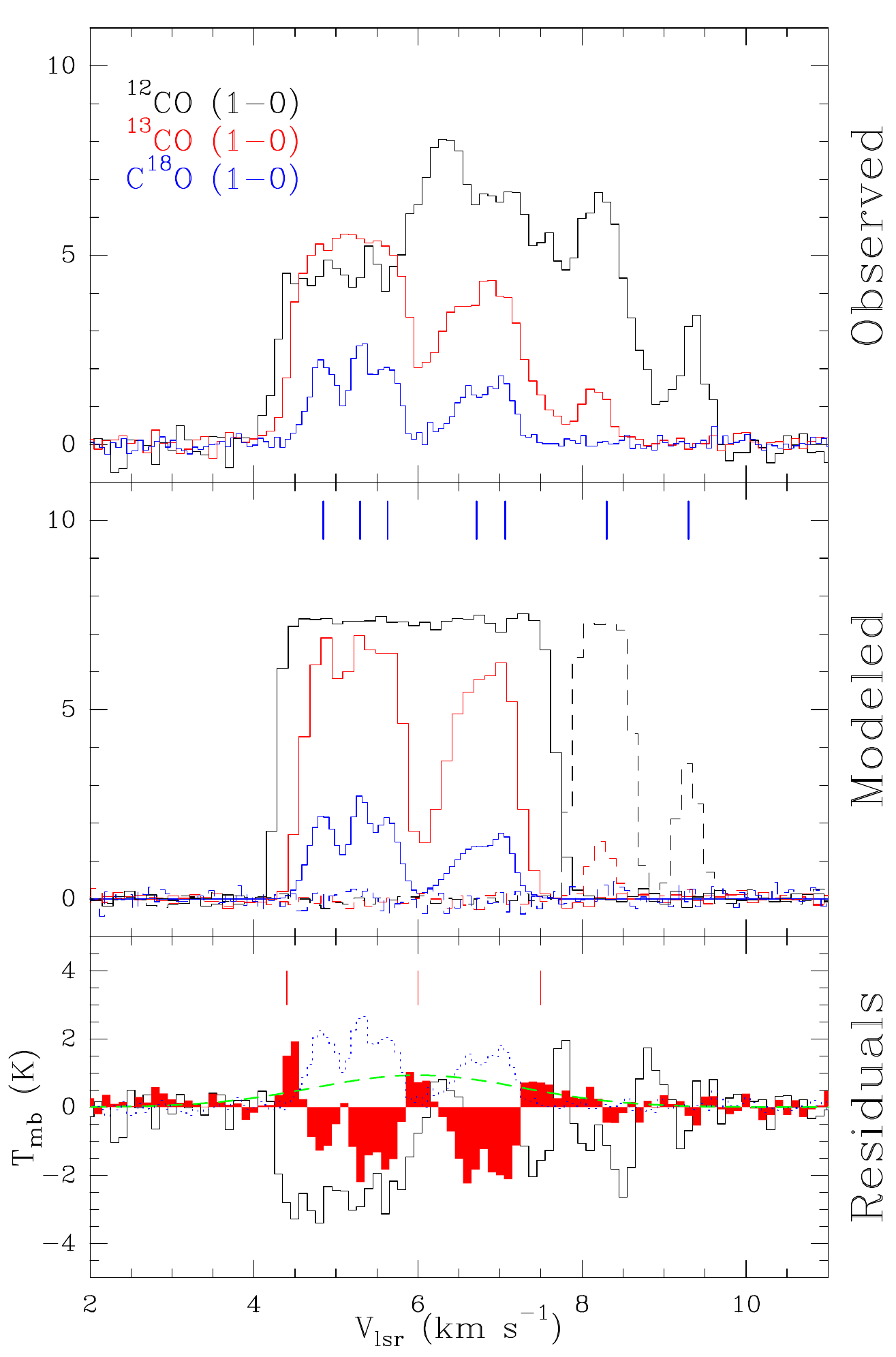}
   \caption{Example of apparently highly supersonic CO spectra in molecular clouds. (Upper panel) Characteristic supersonic (${\cal M}~\sim$~10) spectrum in B213-L1495 observed in C$^{18}$O (1-0) (blue) \citep{HAC13}, and $^{13}$CO (red) \& $^{12}$CO (1-0) (black) lines \citep{GOL08}. (Mid panel) Predicted $^{13}$CO (red solid line) \& $^{12}$CO (1-0) (black solid line) emission from the fitted C$^{18}$O (1-0) line (blue) assuming pure opacity broadening (see text). Within the velocity range where the C$^{18}$O is detected, its saturated emission reproduces most of the observed linewidths of both $^{13}$CO and $^{12}$CO lines. In the case of the two additional components detected in the most abundant $^{13}$CO and $^{12}$CO isotopologues, the expected C$^{18}$O emission is consistent with noise (dashed lines). The central velocity of the 7 velocity components detected in the spectra are indicated by vertical blue lines at the top of the middle panel.
(Lower panel) $^{12}$CO (black line) and $^{13}$CO (red histogram) emission residuals after the subtraction of the corresponding saturated C$^{18}$O emission. In this last case, the observed C$^{18}$O (1-0) emission (blue dotted line) is superposed for illustrative purposes. We note how our simplified LTE treatment overestimates the self-absorption of the optically thick $^{13}$CO and, particularly, the $^{12}$CO lines. The remaining emission found between some of the modeled components observed in the $^{13}$CO wings (marked by vertical red lines in the Lower panel) might suggest the presence of a diffuse substratum (i.e., inter-fiber medium; green dashed line; see Sect.~\ref{sec:intrafiber}). The contribution of gaussian noise with rms=0.2~K was included in all modeled spectra.
}
              \label{fig:complex_lines}%
    \end{figure}

The results in \object{L1517} illustrate how a narrow, sonic-like velocity component detected in optically-thin C$^{18}$O lines could produce apparently broad, transonic linewidths (i.e., ${\cal M \le}$~2-3) in their saturated $^{13}$CO and $^{12}$CO counterparts. However, these opacity effects alone are not able to explain the broad, highly supersonic linewidths with $\Delta V=2-7$~km~s$^{-1}$ (i.e., ${\cal M}\sim$5-15) typically observed in molecular clouds in tracers like $^{12}$CO \citep{SOL87}. To explore the intrinsic linewidths of these broad CO components, in Figure~\ref{fig:complex_lines} (Upper panel) we have studied one of the broadest $^{12}$CO (1-0) spectra in the Taurus molecular cloud \citep{NAR08,GOL08} observed towards the \object{B213}-\object{L1495} region. Interpreted as a single profile, this line presents a total FWHM of $\Delta V\sim4.5$~km~s$^{-1}$ leading to a corresponding non-thermal velocity dispersion of ${\cal M}\gtrsim$~10. Interestingly, this apparently single $^{12}$CO line breaks up into multiple components when the same position is observed in optically thinner lines like $^{13}$CO (1-0) \citep{NAR08} and C$^{18}$O (1-0) \citep{HAC13}. Up to 7 different velocity components are easily identified in our spectra: 5 high-column density components detected in C$^{18}$O plus 2 additional diffuse components only observed in $^{13}$CO and $^{12}$CO. 
The identification of multiple velocity components in Figure~\ref{fig:complex_lines} might suggest that some of the broadest CO lines detected in clouds are actually collections of multiple narrow emission lines. Although less extreme than in the case of L1495-B213, such increase of the line multiplicity observed in the C$^{18}$O spectra in comparison to the more opaque $^{13}$CO and $^{12}$CO lines seems to be characteristic of the CO emission in all molecular clouds and environments as revealed by recent surveys \citep{BUC12,SAD15,WHI15}. 

To test the above hypothesis, in Fig.~\ref{fig:complex_lines} (Mid panel) we simulate the expected saturated $^{13}$CO and $^{12}$CO (1-0) line profiles from the observed C$^{18}$O (1-0) emission in B213-L1495 according to Eq.~\ref{eq:radiative_trans}. The full C$^{18}$O (1-0) line profile was obtained from the gaussian fit of 5 individual velocity components with central velocities of 4.8, 5.3, 5.6, 6.7, and 7.1 km~s$^{-1}$ and transonic, individual linewidths between $\Delta V_{int}\sim 0.3-0.7$ km~s$^{-1}$ (see Table~\ref{table:B213}). At each velocity channel, the C$^{18}$O line opacity derived from Eq.~\ref{eq:tauC18O} is then used to obtained the estimated $^{13}$CO and $^{12}$CO opacities assuming standard fractional abundances plus a constant and homogeneous excitation of T$_{ex}=10$~K (see Sect.~\ref{sec:CoG}). 
Compared to the observed line intensities per channel, the use of the previous gaussian fits allow us to follow the C$^{18}$O emission along the entire velocity window sampled in our spectra. 
This approach permits a first order estimation of the total $^{13}$CO and $^{12}$CO intensities, including those velocity ranges where the C$^{18}$O emission is found within the noise of our spectra (see a discussion on the caveats of these approximations below). 
As a result, and without any additional contribution to the gas velocity field, the modeled $^{13}$CO and $^{12}$CO line profiles mimic the observed total linewidths of these last isotopologues within the velocity ranges where the C$^{18}$O (1-0) emission is detected.  Compared to the $\Delta V_{obs}\sim$~4~km~s$^{-1}$, or ${\cal M}\sim 10$ derived from the direct measurement of the $^{12}$CO FWHM, the intrinsic linewidth of the gas traced in CO is then reduced down to transonic values with ${\cal M}\sim 0.5-2$, that is, a factor of 5 less. 
\emph{Blended in the emission of the most abundant species, the presence of multiple narrow components in combination with optical-depth effects appear as an effective broadening mechanism producing apparently highly supersonic CO lines.}

Although successfully reproducing the observed CO linewidths in Fig.~\ref{fig:complex_lines} (see also Sect.~\ref{sec:diffuse} and Fig.~\ref{fig:complex2}), our oversimplified treatment of the line opacities can only explain the general features observed in the optically thick $^{12}$CO and $^{13}$CO emission. The resulting line profiles are assumed to be toy models of, for instance, more realistic Monte-Carlo modelings \citep[e.g., see similar results obtained in Fig.~9 of][]{TAF98}.
Our predictions should be then taken as a first order approximation of the actual line emission. First, our LTE treatment does not take into account variations of the excitation conditions nor self-absorption effects along the line-of-sight and velocity \citep[e.g., ][]{PIN10}.  Indeed, different gas temperatures and obvious absorption features are already identified towards several of the $^{12}$CO emission peaks detected in Fig.~\ref{fig:complex_lines}.
On the other hand, the assumption of constant isotopic abundances deliberately ignores the detailed treatment of isotopic fractionation effects, selective photodissociation, and ion-molecule exchange occurring at low cloud depths \citep[e.g., ][]{BAL82,VDH88}.
Additionally, our simplified treatment of the line emission does not consider the effects of radiative trapping influencing the optically thick $^{12}$CO line emission at low column densities \citep{EVA99}. 
The combination of these effects could explain the differences between the observed and predicted line intensities in our heavily opaque $^{12}$CO and $^{13}$CO spectra. At least in theory, a detailed NLTE modeling of this line emission and its self-absorption features could be used to obtain additional information of the true cloud structure along the line-of-sight \citep{LEN78}. 

\begin{table}[ht]
\caption{Line emission properties of the gas components detected in Fig.~\ref{fig:complex_lines}.}
\centering
\label{table:B213}
\begin{tabular}{c|c|c|c|c|c}
\hline
\hline
Comp. & $V_{lsr}$  & $\Delta V_{int} $ $^{(1)}$ & $\tau_0$ $^{(2)}$ & $\tau_0 $ $^{(2)}$ & $\tau_0 $ $^{(2)}$  \\
	 & (km~s$^{-1}$) & 	(km~s$^{-1}$) & (C$^{18}$O) & ($^{13}$CO) & ($^{12}$CO) \\
\hline
1	&	4.845	&	0.402	&	0.36	&	2.63	&	201.6	\\
2	&	5.294	&	0.247	&	0.43	&	3.14	&	240.6	\\
3	&	5.630	&	0.335	&	0.34	&	2.48	&	190.6	\\
4	&	6.717	&	0.721	&	0.21	&	1.53	&	117.6	\\
5	&	7.063	&	0.235	&	0.14	&	1.02	&	78.4	\\
6	&	8.300	&	0.350	&	0.04$^{(3)}$	&	0.29	&	22.4	\\
7	&	9.300	&	0.300	&	0.001$^{(3)}$	&	0.007	&	0.56	\\
\end{tabular}
\tablefoot{ $^{(1)}$ Opacity corrected intrinsic linewidths according to Eq.~\ref{eq:broad}. $^{(2)}$ Line opacities derived from Eq.\ref{eq:tauC18O}.
$^{(3)}$ Approximate C$^{18}$O line properties inferred from the $^{12}$CO and $^{13}$CO emission in the spectrum. }
\end{table}%

\subsection{Diffuse molecular gas: transonic CO linewidths}\label{sec:diffuse}

In addition to the high column density gas already traced in C$^{18}$O in Fig.~\ref{fig:complex_lines}, two diffuse gas components are detected only in the$^{13}$CO and $^{12}$CO lines at velocities of approximately 8.2 and 9.3 km~s$^{-1}$. For the 8.2 km~s$^{-1}$ component, the line ratio between the two CO peaks is consistent with central line opacities of $\tau(^{13}CO)\sim$~0.3 and $\tau(^{12}CO)\sim$~22. Upper limits of $\tau(^{13}CO)\lesssim$~0.05 and $\tau(^{12}CO)\lesssim$~3.8 are derived from the line emission at the velocity of 9.3 km~s$^{-1}$. In both cases, the expected C$^{18}$O intensities for these two lines are all within the noise level of our spectra. Remarkably, once the opacity effects are subtracted in these two cases their intrinsic linewidths of these diffuse components are similar to the typical linewidths detected in those C$^{18}$O rich components (see Table~\ref{table:B213}).

   \begin{figure}[ht!]
   \centering
   \includegraphics[width=\columnwidth]{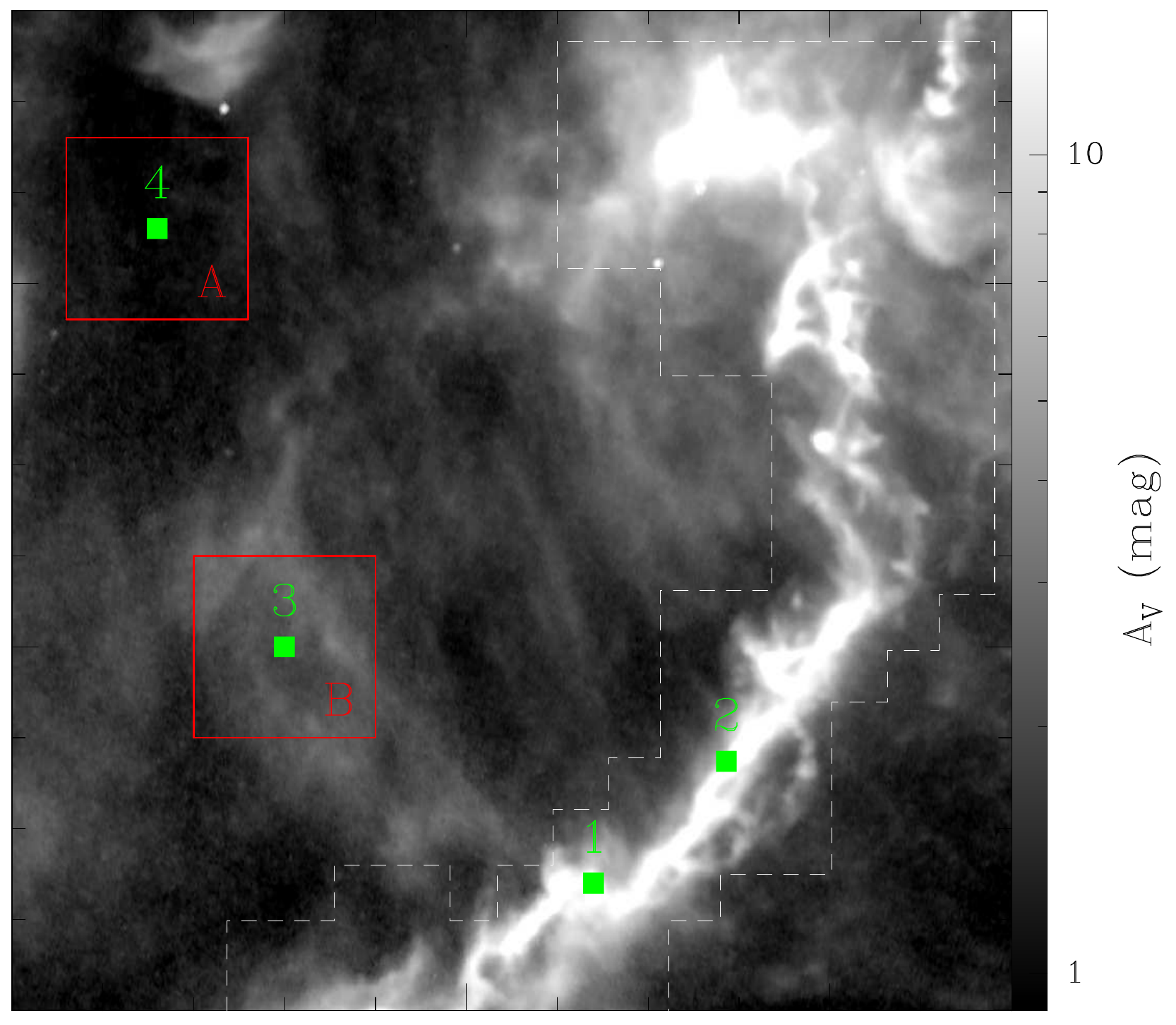}
   \caption{Herschel-continuum total column density map of the northern part of B213-L1495 \citep{PAL13}. The four positions presented in Fig.\ref{fig:complex2} are indicated in the map. Note the presence of several prominent large scale striations at low column densities. The red squares indicate the two sections studied in detail corresponding to representative positions of the Mask 1 (Area A) and Mask 2 (Area B) regions in \citet{GOL08}, respectively. The white dashed-line encloses the high-column density, C$^{18}$O-bright region surveyed by \citet{HAC13}.}
              \label{fig:B213_map}%
    \end{figure}

   \begin{figure*}[ht!]
   \centering
   \includegraphics[width=\textwidth]{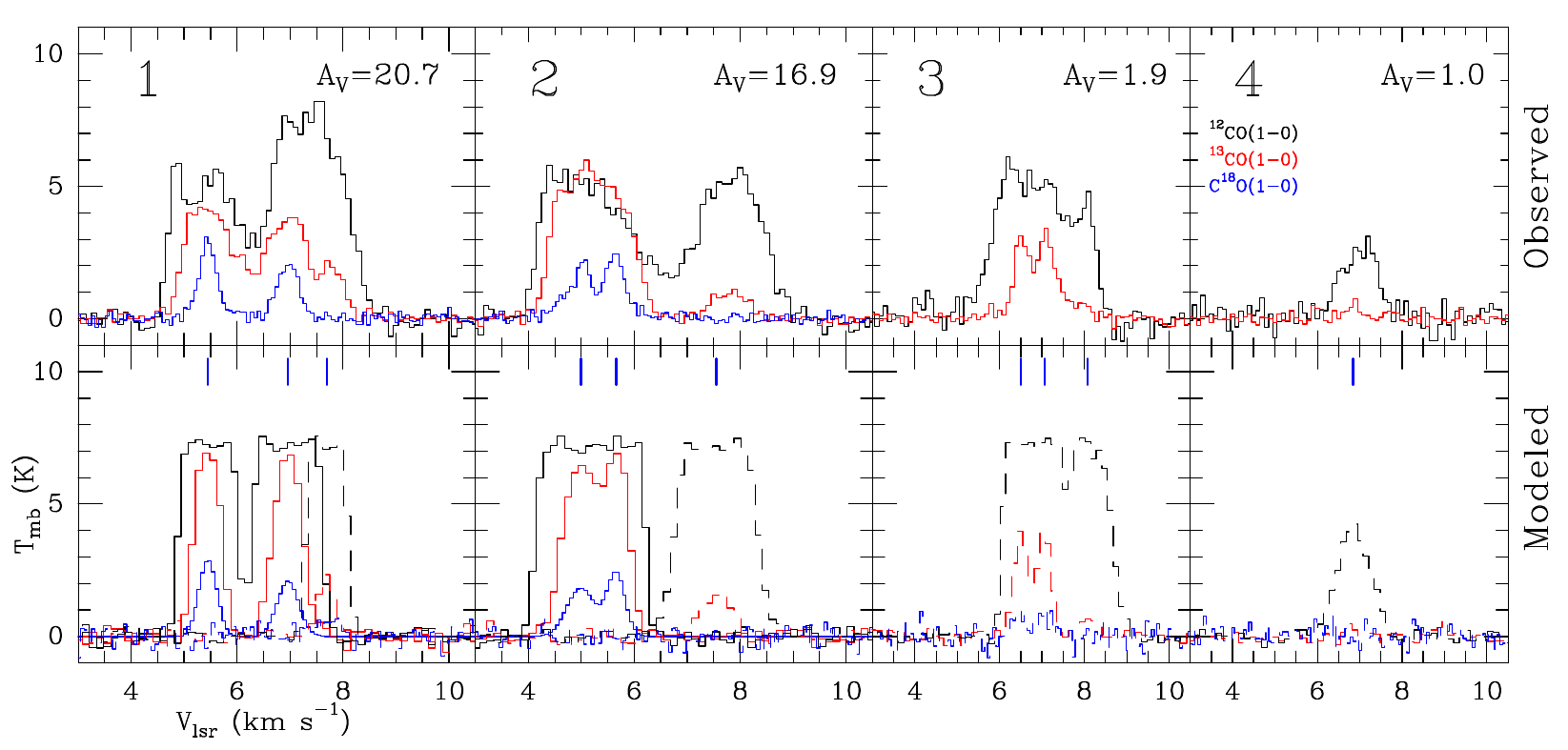}
   \caption{Observed (Upper panel) and Modeled (Lower panel) C$^{18}$O (1-0) (blue), and $^{13}$CO (red) \& $^{12}$CO (1-0) (black) line emission the B213-L1495 region (positions 1 to 4). The location of these spectra are indicated in the map on Fig.~\ref{fig:B213_map}. The total  visual extinction of gas at these positions, derived from Herschel continuum observations, are indicated in the upper right corner. Multiple peaks are identified in the $^{13}$CO and $^{12}$CO spectra when these lines become optically thin (e.g., position 3). Symbols and lines similar to those in Fig.~\ref{fig:complex_lines}. }
              \label{fig:complex2}%
    \end{figure*}

It is commonly presumed that the low column density gas traced in $^{12}$CO is represented by broad and, therefore, highly dynamic linewidths. 
Due to their special location in the B213-L1495 region, the two diffuse components detected in Fig.~\ref{fig:complex_lines} might not be considered as representative of the diffuse gas observed in the whole Taurus molecular cloud.
Also, and still not completely ruled out by our method, some of the opaque lines coexisting within this regions could potentially hide a much broader and supersonic gas component (see also Sect.~\ref{sec:intrafiber}).
However, their unexpectedly narrow velocity dispersions raise a question: \emph{What is the actual intrinsic velocity dispersion of the most diffuse gas components traced in $^{12}$CO?}
To explore this issue, in Fig.~\ref{fig:complex2} we compare the emission of the $^{12}$CO, $^{13}$CO, and (when possible) C$^{18}$O (1-0) lines in 4 representative locations around the B213-L1495 cloud. These positions cover a wide rage of column densities, both along this filament (positions 1 and 2; A$_V >$~15 mag) and its diffuse environment (positions 3 and 4; A$_V < $~1-2 mag). Aiming to observe optically thinner lines, the last two positions were restricted to positions at angular distances of $\sim$~0.5~deg (i.e., >~1~pc) away from any high-column density structure with A$_V >$~5 mag) detected in any previous C$^{18}$O \citep{ONI96}, extinction \citep{LOM10,SCH10}, or Herschel-continuum \citep{PAL13} measurements. 
In all cases, the now reduced line opacity in these diffuse components reveals the otherwise saturated line substructure of the CO emission. As in the case of Fig.~\ref{fig:complex_lines}, the most extreme and wide $^{12}$CO line profiles can be directly explained by the combined effect of optically thick effects and multiple components with intrinsic narrow $\Delta V_{int}\sim 0.35-0.80$ km~s$^{-1}$. Remarkably low velocity dispersions are also recovered in the observed$^{12}$CO (1-0) lines at low column densities when this typically highly optically thick line has only moderate optical depth due to low column density (see position 4 in Fig.~\ref{fig:complex2}), in agreement with our predictions.

We have extended our analysis by studying the velocity dispersion of the low- column density gas detected only in $^{12}$CO (1-0) (subregion A) and the $^{13}$CO plus $^{12}$CO (1-0) lines simultaneously (subregion B) around the B213-L1495 filament (see Fig.~\ref{fig:B213_map}). These two regions show equivalent averaged extinction values of A$_V=1.1\pm0.1$ (or N$(H_2)~\sim~10^{21}$~cm$^{-2}$) and A$_V= 2.2\pm0.4$ (or N$(H_2)~\sim~2\times 10^{21}$~cm$^{-2}$), respectively, and are representative of the Masks 1 and 2 defined in \citet{GOL08}. To improve the spectra sensitivity required for our analysis, we have convolved and regridded the original \citet{GOL08} $^{12}$CO and $^{13}$CO FCRAO datasets into a new Nyquist sampled map with a final resolution of 60 arcsec. The surveyed areas consist of a final sample with more than 1100 spectra within a 1000 by 1000 arcsec$^2$ each. Similar to previous Sections, we have investigated the observed line profiles of the, in each case, thinnest CO isotopologue by fitting gaussians within the observed subregion. This gaussian decomposition follows the fitting strategy described in \citet{HAC13}.

   \begin{figure}[ht!]
   \centering
   \includegraphics[width=\columnwidth]{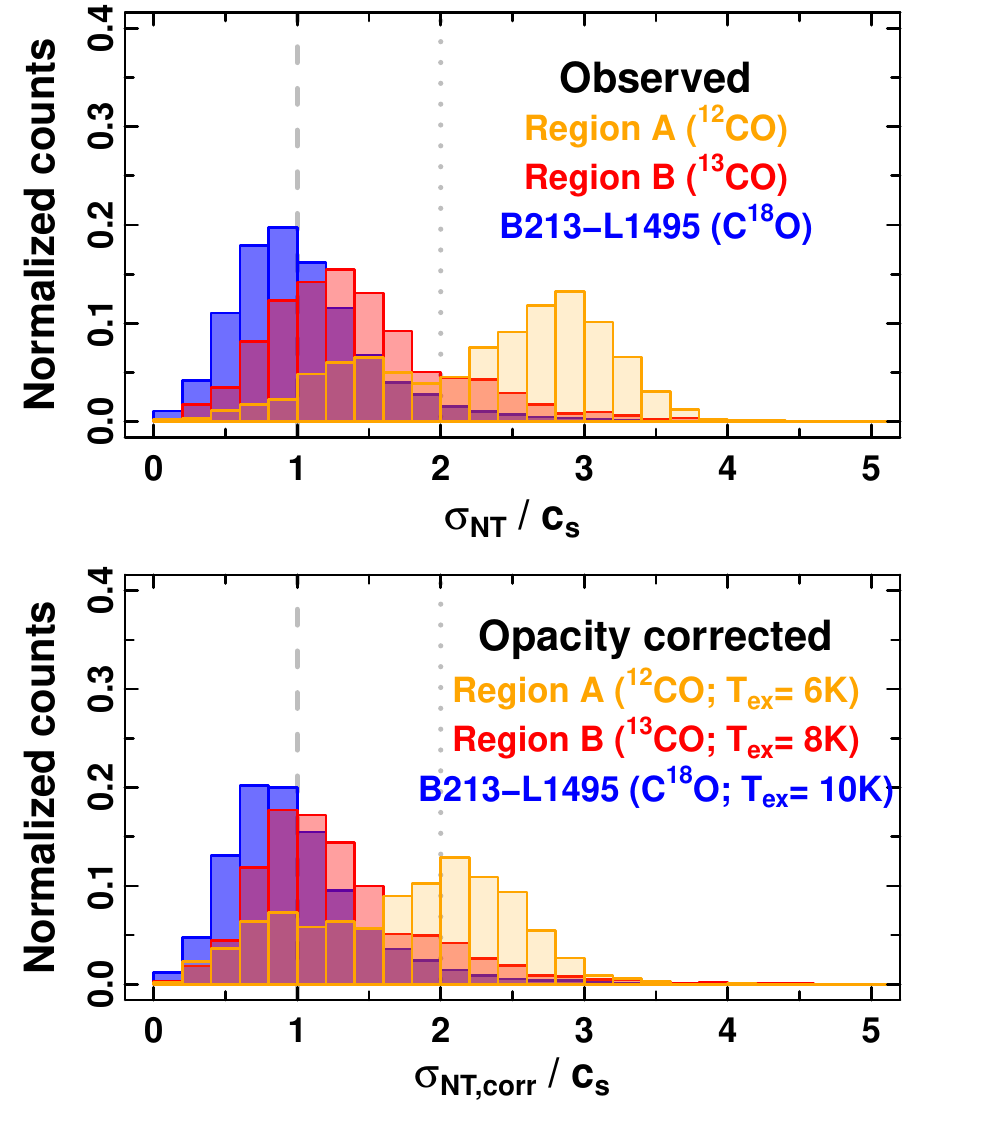}
   \caption{Comparison of the non-thermal velocity dispersions (i.e., $\sigma_{NT}$, in units of the sound speed at 10~K) obtained directly from the observed FWHM (Upper panel) and after corrections of the opacity broadening (Lower panel) for the three regions studied in Fig.~\ref{fig:B213_map}: Region A using $^{12}$CO (A$_V\sim1$; >1100 spectra); Region B using $^{13}$CO (A$_V\sim2$; > 1100 spectra); and the L1495-B213 filament using C$^{18}$O (A$_V>5$; 23\,000 spectra \citep{HAC13}). The opacity corrected values of $\sigma_{NT,corr}$ were obtained from the intrinsic linewidths $\Delta V_{int}$ after correcting the observed linewidths using Eqs.~\ref{eq:broad} and \ref{eq:tauC18O}.
   Sub-thermal excitation values (i.e., T$_{ex}$=6 \& 8~K) are used for the opacity correction for those points in Regions A and B, in agreement to \citet{GOL08}. Due to the expected increase of the gas kinetic temperature in the cloud outskirts, the observed values for $\sigma_{NT}$ in Regions A and B correspond to upper limits of the intrinsic velocity dispersion of the gas. Only points with SNR$\le$~3 are considered.}
              \label{fig:diffuse_hist}%
    \end{figure}

In Figure~\ref{fig:diffuse_hist}, we represent the distribution of both the observed (Upper panel) and the opacity corrected (Lower panel) non-thermal velocity dispersions of the gas in these Region A (from $^{12}$CO) and Region B (measured in $^{13}$CO) in comparison with the results obtained in B213-L1495 by \citet{HAC13} (i.e., C$^{18}$O). The opacities are estimated in each case using Eq.~\ref{eq:tauC18O} adapted for each of these CO isotopologues assuming excitation temperatures of T$_{ex}=$~6, 8, and 10 K, respectively, in agreement to the mean values derived for the CO isotopologues in the different density regimes (Masks) studied by \citet{GOL08}. Although the gas temperature is expected to rise towards those column density regimes traced in Regions B and A, respectively, we conservatively compare these values with the expected sound speed of the coldest gas in the cloud at 10~K. Even when no opacity correction is applied, the maximum non-thermal velocity dispersions observed in the Taurus cloud are limited to values of $\sigma_{NT}/c_s\lesssim~3$ at column densities of A$_V\gtrsim1$. These calculations likely underestimate both the line opacity and the actual thermal velocity dispersion of the gas in these low column density regions \citep[e.g., see][]{PIN10}. Our values should be then taken as upper limits of the actual velocity dispersion on the gas.

In the light of the above, our preliminary results suggest that most low column density gas detected in the most abundant CO isotopologues  presents maximum intrinsic velocity dispersions consistent with Mach numbers ${\cal M}\lesssim 2-3$ at most. 
When compared with the C$^{18}$O rich components studied in regions like B213-L1495, the net increase measured in the intrinsic velocity dispersion between the low- and high column density material traced in CO would be then limited to a factor of $\lesssim 2$ and restricted to changes within (or close to) the transonic regime.
Interstingly, CO spectra with similarly narrow linewidths (sometimes also associated to multiple components) are also identified at low column densities in both high-latitude and translucent clouds \citep{KET86,VDH91}. 
The similarities with the line profiles identified in regions of higher column density and associated with the presence of fibers (Sect.~\ref{sec:supersonic}) suggest a similar organization of the gas structure at lower column densities. These hypothesis is reinforced by the detection of multiple velocity-coherent structures at low column densities in Taurus \citep{PAN14}, the filamentary nature of the low-column density striations dominating the diffuse CO emission \citep{GOL08,PAL13},  and the spider-web appearance of diffuse and high-latitute clouds like Polaris \citep{AND10}. 
Deeper observations of diffuse regions and detailed analysis of the different CO emission profiles are required to further explore these latest conclusions\footnote{We remark here that the transonic nature of the CO emission pointed out above is referred to the pristine dynamical properties of the molecular gas traced by these lines. This description does not rule out the localized presence of high-velocity line wings and dissipative structures \citep{FAL90} or the truly supersonic lines detected in outflows and/or stellar winds \citep[e.g., ][]{KWA76,SNE80} }.


\section{Discussion: Macroscopic turbulence and Tangled Molecular Clouds}\label{sec:discussion}

Two main conclusions can be drawn from our analysis of the observed CO linewidths presented in Sections \ref{sec:L1517} to \ref{sec:diffuse}. First, the individual $^{12}$CO and $^{13}$CO lines are systematically affected by opacity broadening effects resulting in apparently broad, suprathermal lines. After the correction of these optically thick effects, the intrinsic velocity dispersions of the gas traced by these two CO isotopologues approach the (tran-)sonic lines observed in less abundant, and therefore less opaque, tracers like C$^{18}$O. Second, line blending effects favored in optically thick regimes can lead to several km~s$^{-1}$ wide $^{12}$CO lines. Rather than a single broad component, the detection of multiple gaussian-like peaks in the optically-thin $^{13}$CO and C$^{18}$O lines confirm the presence of multiple quiescent velocity components superposed along the line-of-sight. This superposition of velocity components could particularly affect surveys of unresolved objects (e.g., clouds or galaxies) both in local and extragalactic contexts. This underlying fine substructure appears as a fundamental characteristic of the internal gas kinematic in molecular clouds. 

\begin{figure*}[ht!]
   \centering
     \includegraphics[width=\textwidth]{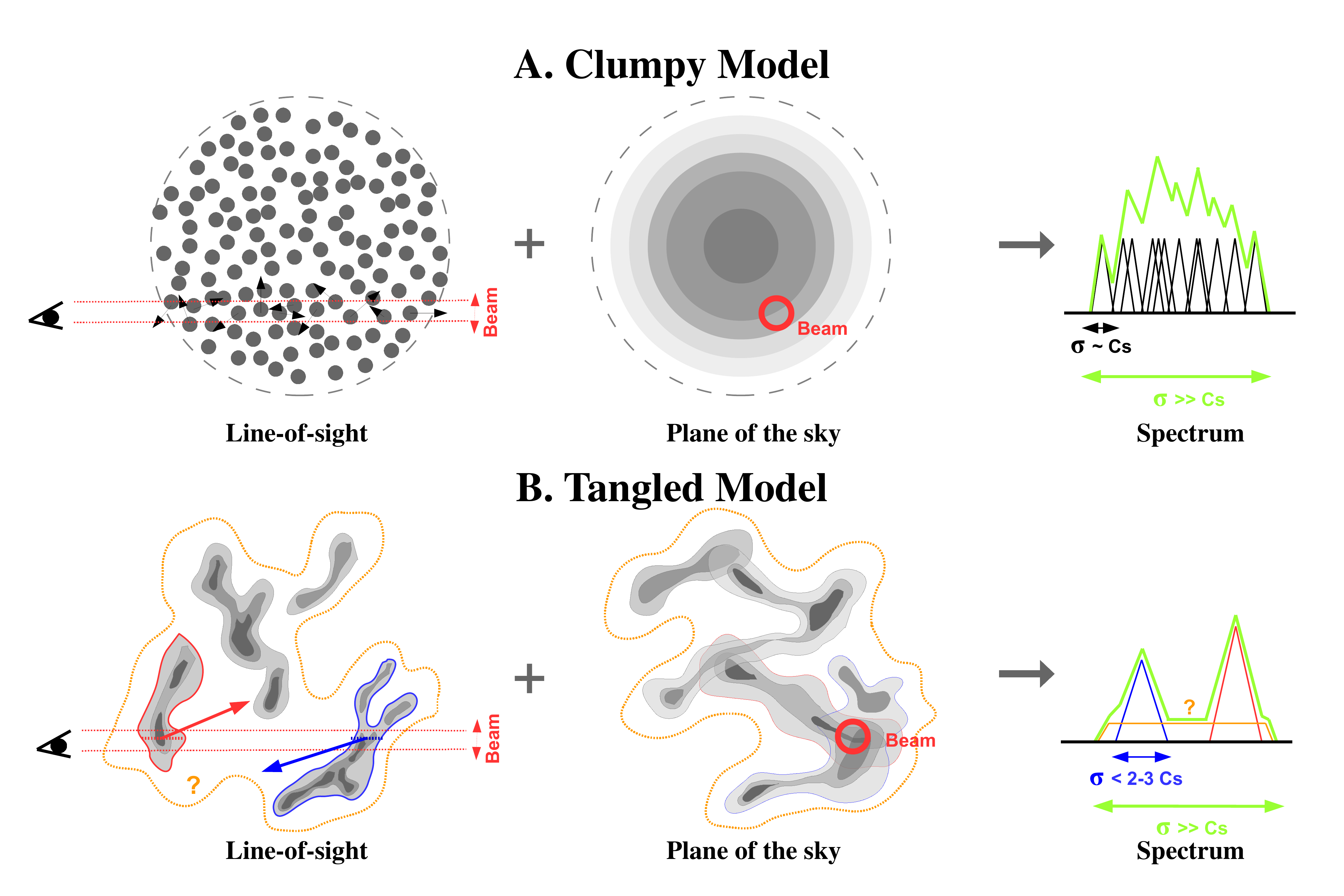}
   \caption{Illustrative cartoon for the Clumpy (Upper panel; inspired by Fig.~4 in \citet{WOL93}) and Tangled (Lower panel) macrotubulent models. The sketches represent the corresponding line-of-sight (Left) and projected (Center) views of the cloud, as well as the expected line profiles in the case of optically thin emission (Right; e.g., C$^{18}$O observations). Note how the broadest and highly supersonic linewidths have different origins in these two models. Due to their large filling factor along the line-of-sight, they are generated by the accumulation of a high number of sonic-like clumps at different velocities in the classical clumpy models. On the contrary, in the proposed tangled clouds, the linewidths arise from the superposition of few transonic structures correlated at large scales, both spatially and in velocities (i.e., fibers).}
              \label{fig:cartoon}%
    \end{figure*}

The above results qualitatively resemble the early called``macro turbulent'' models for the ISM introduced after the first CO observations in molecular clouds \citep{KWA86,SOL87,ROB93}. 
Instead of the continued density and velocity fields assumed in the microturbulence framework, 
in models dominated by macroscopic turbulence the internal structure of molecular clouds is created by multiple and discrete small-scale, thermal-like (i.e., $\sigma_{int}\sim \sigma_{th}$) clumps moving ballistically with large interclump velocities (i.e., $\delta V_{lsr}>\sigma_{th}$). Within the observed beam sizes, these macroturbulent models classically assume a high filling factor clumpy medium created by a large number ($\gg$~1) of spatially unresolved cloudlets ($\ll$~0.01~pc) superposed along the line-of-sight \citep[e.g., ][; see a cartoon in Fig.~\ref{fig:cartoon}, Upper panel]{TAU91,FAL91}. 

Radiative transfer calculations have proven the intrinsic differences between CO emission properties and line shapes arising from these micro- and macroscopic media \citep{BAK76,LEN78,WOL93}. In line transfer problems, a distinction between these models can be physically quantified by the ratio between the photon mean free-path $\tau$ and the correlation length $L$ of the gas velocity field (i.e., $\sim\frac{L}{\tau}$): {\it short correlation lengths are expected for media dominated by microscopic motions while large spatial correlations in velocity are characteristic of fluids dominated by macroscopic turbulence} \citep[see][for a detailed discussion]{LEN78}.
As result, clouds ruled by microturbulent motions tend to produce CO lines with flat-topped profiles and strong self-absorption features. In contrast, the relative and discontinuous motions between individual cloudlets in macroturbulent models lead to CO line profiles presenting discrete and multiple emission peaks in agreement to our observations.

The above differences describe the type of turbulent motions inside molecular clouds depending on whether they are observationally resolved (macro-turbulence) or unresolved (micro-turbulence). As important point is that this distinction is independent of the scales where these motions are driven (i.e., injected). Due to the multi-scale nature of ISM turbulence \citep[e.g.,][]{MAC04}, both macro- and microturbulent regimes can be simultaneously found in either large-scale (LSD) or small-scale (SSD) driving turbulence scenarios \citep[e.g., see examples of both LSD and SSD models in][]{KLE00}. The characterization of these two independent parameters has been previously attempted using both line profiles (type: macro- vs. micro-turbulence) and the energy power spectrum (driving scale: LSD vs. SSD) from different analysis techniques like velocity differences and momenta, structure functions, $\Delta$-variance, and velocity PDFs \citep[][; see also Sect.~\ref{sec:KinEne}]{MIE94,OSS02,HAC15}.

Noticeable differences are pointed out in our observations compared with most of the previous macroturbulent models.
The analysis of our CO lines in Taurus suggests a reduced filling factor (see also Sect.~\ref{sec:intrafiber}), typically limited to < 10 structures per beam, at both low- and high column density regimes. From the analysis of the Position-Position-Velocity space in the case C$^{18}$O emission in B213-L1595, \citet{HAC13} have pointed out the spatial correlation of the velocity components detected in these spectra at scales of $\gtrsim$~0.5~pc, that is, much larger than the physical resolution of the current radiotelescopes in nearby molecular clouds. By the reconstruction of their emission in the Position-Position-Velocity space, these authors also demonstrated that the presence of multiple peaks detected in the high-column density C$^{18}$O spectra are generated from the combination of multiple velocity-coherent filamentary-like structures, named as fibers.
Although not necessary elongated, the narrow well-defined CO components detected in both low- and high- column density regions suggest a similar superposition of discrete and fully-resolved gas structures inside clouds at different densities. 
In this context, a generalization of the term \emph{fiber} would be referred to the individual cloudlets highly correlated at large-scales both spatially and in velocity forming a molecular cloud, irrespectively to their geometrical shape. 
Hundreds of these fiber-like structures have been identified in the analysis of the $^{13}$CO emission in the whole Taurus cloud \citep{PAN14}.

Molecular clouds that consist of large-scale fiber-like, elongated structures arise naturally in numerical simulations of diffuse, warm HI gas
flows that undergo a cooling instability, coupled with various hydrodynamical instabilities \citep{BAL99,HEI08a,NTO11}. Another possibility might be gravitational modes, sweeping up material at the edges of collapsing structures \citep{BUR04,HAR07,VAZ07,HEI08b}. The situation might be even more complex when taking into account the galactic environment where molecular clouds might form and grow by interactions of filamentary, already partly molecular dense gas structures \citep{DOB12,DOB14}.
Similarly, simulations have also reported the widespread presence of a complex and intrincated substructure in molecular clouds both spatially and in velocity \citep[e.g.,][]{MAC00,OSS01,OSS02}. Pre-existing diffuse fibers prior to the formation of massive and high column density regions have been recognized in simulations by \citet{MOC14,SMI14,SMI16}.
These structures might be linked to the intrinsic intermittent nature of turbulence \citep{FAL91} again understood as a process at cloud scales.
\emph{Nested complex networks of a large but still measurable number of these fibers would then create the appearance of a tangled substructure inside molecular clouds} (see a cartoon in Fig.~\ref{fig:cartoon}; Lower panel).

This updated description of the turbulent motions inside molecular clouds explains the emission properties observed of the different CO tracers.  
Each individual fiber (or cloudlet) appears to be internally dominated by (tran-)sonic ($\cal{M}\lesssim$~2), unresolved, and isotropic velocity dispersions created by random internal motions (i.e., microturbulence).
On the other hand, their supersonic motions ($\cal{M}>$~2) are created by the relatively large velocity differences between these substructures (aka macroturbulence). These fully-resolved intraclump velocities are referred to highly anisotropic, systematic motions ordered at large scales (i.e., $\gg$ beam size in nearby clouds). As demonstrated in previous Sections, this underlying discretized substructure can be easily misinterpreted as part of the unresolved (micro-)turbulent dynamics in observations of the most abundant CO isotopologues due to the likely combination of these individual entities in large number (e.g., when all the gas velocity components forming a cloud are diluted within a single beam in extragalactic studies) in addition to the line blending (Sect.~\ref{sec:supersonic}) and optical broadening effects (Sect.~\ref{sec:L1517}). 

In the following sections we will explore some observational characteristics of these fully-resolved, macroscopic motions inside molecular clouds.
Although particularly referred to these fibers, and due to the limited coverage of our observations, we restrict our study to the statistical properties of a medium created by a superposition of multiple independent entities. We remark that this description is based on quantified observables measured from the molecular gas traced in CO. Besides the tangled structure of molecular clouds, the validity of most of the assumptions and analysis presented below should, however, hold for any clumpy medium producing similar spectra to those presented in this work. 

\subsection{Interpretation of the CO emission}

The various CO isotopologues are usually misinterpreted as density selective gas tracers. It is presumed that while $^{12}$CO and $^{13}$CO trace the most diffuse gas in the cloud, only molecules like C$^{18}$O and C$^{17}$O are sensitive to the densest and deeply embedded gas regions. This concept would in principle appear to be confirmed by the detection of the most diffuse gas regions in only the two most common CO isotopologues, but not in their rarer counterparts. 
However, the similar physical and chemical properties of all these isotopologues (energy levels, Einstein coefficients, depletion...) make their excitation conditions quasi-identical. Only radiative trapping effects in the highly opaque $^{12}$CO lines appears as differentiating process favoring the excitation of these transitions at low gas densities.
If both photodissociation and isotopic fractionation are neglected, the observed differences are mainly created by the large abundance ratios between these molecules.   
Sensitive to almost the same gas regimes, most of the observationally biased low signal-to-noise spectra are responsible for the non-detection of the C$^{18}$O and C$^{17}$O in low column density regions\footnote{For instance, the detection of an optically thin $^{12}$CO (1-0) line with T$_{mb}=1$~K would lead into an equivalent C$^{18}$O (1-0) emission with T$_{mb}\sim1/560$~K under LTE conditions. An observation aiming to detect such weak line would then require a total integration time of a factor $560^2$ longer in order to detect this last transition with similar SNR than $^{12}$CO.}. 

Although tracing the same gas, the large relative abundances of the $^{12}$CO and $^{13}$CO isotopologues in comparison to C$^{18}$O inevitably translate into larger opacities. As discussed in Sect.~\ref{sec:CoG}, the frequency dependent opacity effects mostly affect the bulk of the line emission at its central velocities. While limiting the study of the total column density of the gas, their relatively low opacity line wings still reflect most of the gas kinematics of the gas traced in CO. 

These conclusions are supported by our observational results. In all the position where the three main CO isotopologues are detected in our data, the subtraction of the opacity broadening effects in the $^{12}$CO and $^{13}$CO lines make their kinematic properties and total intensities almost indistinguishable from those of C$^{18}$O (Sect.~\ref{sec:L1517} and \ref{sec:supersonic}). As seen in Fig.~\ref{fig:complex_lines} (Lower panel), comparisons between our observations and models show that the LTE-saturated versions of the C$^{18}$O emission reproduce more than $\sim$~90\% of the total linewidth of the more abundant $^{12}$CO and $^{13}$CO lines. Most of (if not all) the gas kinematics and column density information of the molecular gas is consistent with the parameters already derived from the optically thin C$^{18}$O molecule. Only when the C$^{18}$O emission drops below the detection thresholds other tracers are required. In such cases, the now optically thin $^{13}$CO and even  $^{12}$CO lines recover the same kinematic properties of the C$^{18}$O gas (Sect.~\ref{sec:diffuse}).

\subsection{Filling factor and Inter-fiber medium}\label{sec:intrafiber}

Classical clumpy models assume clouds to be formed by a large number of small, unresolved clumps. An estimate of the minimum number of clumps superposed along the line-of-sight can be obtained from the ratio between between the typical linewidth of each of these clumps ($\Delta V_{clump}\sim0.1$ km~s$^{-1}$; e.g., see \citet{WOL93}) and the total observed linewidth ($\Delta V$) as $ \mathrm{n}_{clumps}\ge\frac{\Delta V}{\Delta V_{clump}}$, assuming a minimum overlap in velocity between clumps (see Fig.~\ref{fig:cartoon}, Upper panel). At each observed position, the total column density N$(\mathrm{H}_{2})$ is proportional to the number of clumps superposed in velocity along the line-of-sight: 
\begin{equation}\label{eq:Nclumps}
	\mathrm{N}(\mathrm{H}_{2})=\sum_i^{\mathrm{n}_{clumps}} \mathrm{N} _i \simeq \mathrm{n}_{clumps} \times \langle \mathrm{N}_{clump} \rangle
\end{equation}
Densely populated (i.e., $ \mathrm{n}_{clumps}\gg 1$), this clumpy configuration then leads to large filling factors per beam $f|_{clumpy}$ 
and compact clumps with low individual column densities N$_{clump}$. For the same cloud depth along the line-of-sight, the low number, intensity, and fully-resolved peaks detected in our C$^{18}$O spectra along molecular clouds like Taurus suggest a much lower filling factor ($f|_{tangled} < f|_{clumpy}$) created by fewer (n$_{fibers}\ll\mathrm{n}_{clumps}$) but comparably higher column density structures (i.e., N$_{fiber} \gg $~N$_{clump}$) (see schematic cartoon in Fig.~\ref{fig:cartoon}).

Approximately $\lesssim$~10\% of total integrated emission detected in the high abundant tracers like $^{12}$CO and $^{13}$CO remains unexplained by our LTE extrapolation of the C$^{18}$O lines. This emission excess is particularly visible in the $^{13}$CO emission in spectra like the one presented in Fig.~\ref{fig:complex_lines} (see Lower panel). The preferred location of this excess emission at the line wings seems to dismiss additional high column density components not detected at the current C$^{18}$O sensitivities.
As discussed in simple regions like L1517 (Sect.~\ref{sec:L1517}), this remaining emission can be explained if the abundance of this last molecule were underestimated by a factor of two or if its excitation were lower than the expected 10~K for the gas kinetic temperature. However, in more complex regions like B213-L1495 the emission from a diffuse stratum can not be ruled out. This inter-fiber medium would correspond to some diffuse material encompassing some of the denser structures in the cloud. Masked by other optically thick CO components, its emission might still be detectable at the lower-opacity line wings in our spectra (see also Fig.~\ref{fig:cartoon}). 

While its existence can not be confirmed with the data at hand, several properties of this inter-fiber molecular gas can still be constrained from our observations.
First, its emission is primary detected at the same velocities as the most prominent gas components detected in C$^{18}$O. If it exists, this medium most likely would permeate the high  density fibers in the cloud. Second, its total column density should be restricted to equivalent extinction values of A$_V\lesssim$~1, otherwise its emission would be detected in our high signal-to-noise $^{12}$CO and $^{13}$CO spectra at low extinctions (see Fig.~\ref{fig:complex2}). Third, extended between fibers, its density would be of the order of a few hundreds per cubic centimeter. The low excitation levels produced in such low densities would produce low intensity lines, in agreement to our spectra. Finally, and related to these last properties, its emission might be therefore highly diluted in velocity. Only this inter-fiber medium could actually present highly supersonic (i.e., several km~s$^{-1}$) velocity dispersions.

Widely extended and at low column densities, this diffuse inter-fiber material seems to be of secondary importance for the evolution of the cloud. Although some emission can be hidden in lines with low excitation conditions, the comparison of its emission with the prominent and well defined gas components detected in C$^{18}$O suggests that most of the molecular gas detected in CO is actually concentrated in higher density fibers rather than this more diffuse inter-fiber gas. Because of its low density, this inter-fiber medium would also be dynamically unimportant. Orders of magnitude lower in density, its interaction with the denser fiber-like structures should be negligible. However, some of these fibers could still gain mass while moving through (and sweeping up) this medium \citep{TAF15}. 

Depending on its kinetic temperature, the inter-fiber medium could nevertheless provide some additional thermal pressure. The presence of large amounts of warm atomic gas inside molecular clouds seem to be ruled out according to previous HI self-absorption observations in these objects \citep{LI03}. On the other hand, theoretical models \citep{WOL10} predict that $\sim$~30\% of the total molecular mass in clouds resides in the so-called CO-dark molecular gas \citep{GRE05,PIN13}. Found at the cloud edges, this gas component is characterized by its low (or inexistent) emission in CO and H$_2$ rotational transitions due to the combination of low excitation conditions and CO photodissociation. Describing analogous conditions to the CO-dark gas, \citet{GOL13} reported values of n(H$_2$)~=~68-92~cm$^{-3}$, T$_K=~50-100$~K, and P/k~=~4600-6800~K for the gas content in diffuse molecular clouds. The low filling factor in our tangled clouds would facilitate the diffuse interstellar radiation field to effectively permeate the innermost parts of the cloud creating the conditions needed to produce this warm molecular medium. In this speculative scenario, a similar inter-fiber medium could help to pressure confine some of the densest CO components detected in our spectra. 
This inter-fiber medium could also participate on the gravitational collapse of the cloud at large scales or shaping the cloud at large scales.

\subsection{Column Density Distribution: Superpositions and component evolution}\label{sec:ColDen}

If a medium is formed by the superposition of multiple, independent fibers, and assuming that each of these structures have a characteristic column density $\langle\mathrm{N}_{fiber}\rangle$, it is then obvious that the projected total column density of the cloud at a given point should linearly increase with the number of structures along the line-of-sight (or superpositions per beam) according to Eq.~\ref{eq:Nclumps}. A systematic increment of the observed number of components (i.e., peaks per spectrum, N$_{peaks}$) is then predicted at higher column densities as $\langle \mathrm{N}(\mathrm{H}_2)\rangle \sim \mathrm{N}_{peaks} \times \langle\mathrm{N}_{fiber}\rangle$.  

\begin{figure}[h!]
   \centering
     \includegraphics[width=\columnwidth]{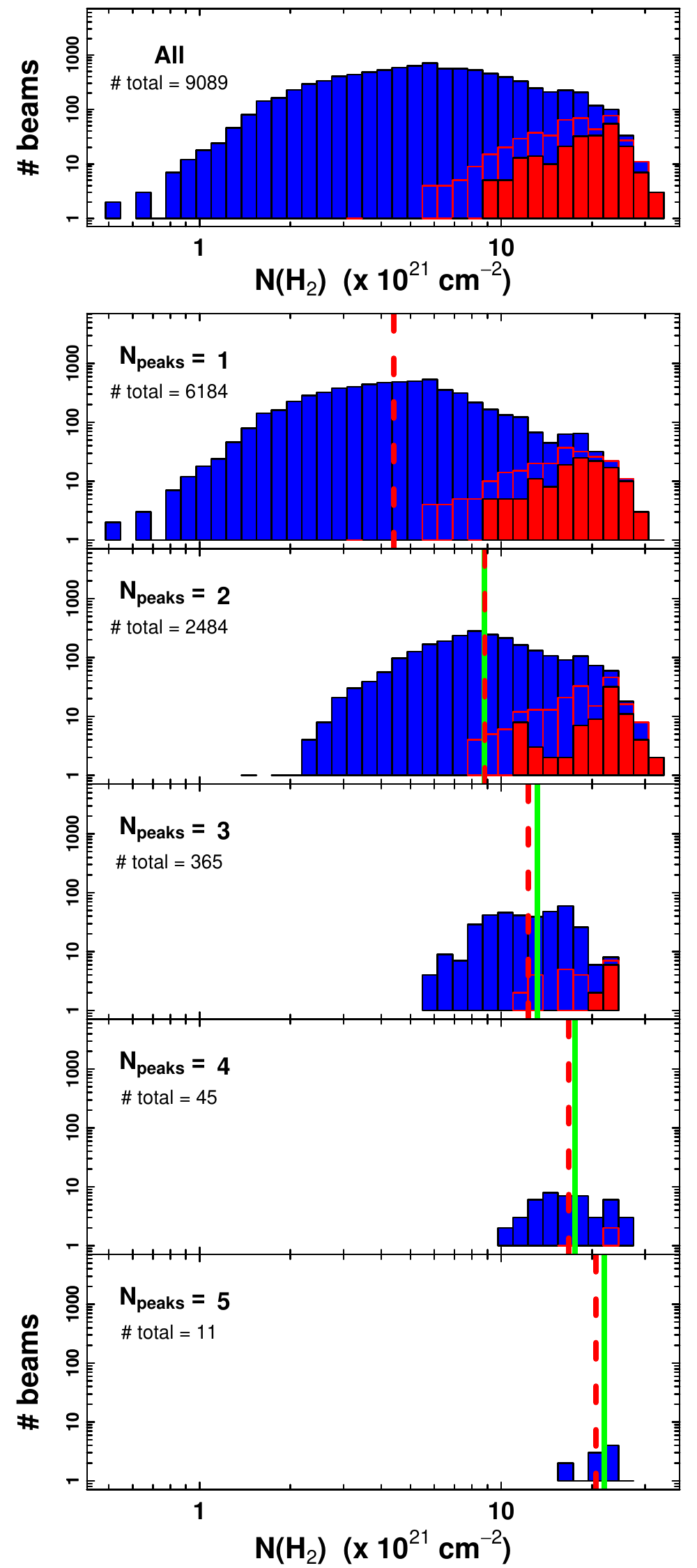}
   \caption{Distributions in log-log space of the total column densities derived from the C$^{18}$O observations (blue) in the B213-L1495 by \citet{HAC13}: (Upper panel) Total column densities for all the positions fitted with at least one component with SNR~$\ge$~3; (Lower subpanels) Corresponding column density distributions for positions with a number of N$_{peaks}=$~1 to 5. Those positions containing at least one dense component (SNR$(N_2H^+)\ge$~3; red line bars) and cores (SNR$(N_2H^+)\ge$~6; solid-red bars) are indicated in histograms in the corresponding plots.}
              \label{fig:superpos_histo}%
    \end{figure}

We have investigated the correlation between the total column density and the line multiplicity in the B213-L1495 region from the results presented by \citet{HAC13}. For each C$^{18}$O spectrum fitted with at least one component with SNR~$\ge$~3, the total column density per beam is obtained from the integrated emission of all the components detected in this molecule assuming LTE excitation conditions at 10~K and standard C$^{18}$O abundances \citep{FRE82}. Simultaneous SNR~$\ge$~3 detections of coexisting N$_2$H$^+$ (1-0) components are employed to identify those spectra containing dense gas ($n(H_2)>10^4$~cm$^{-3}$). Among them, the gas belonging to the 19 dense cores identified within this region are selected as those positions with SNR$(N_2H^+)\ge$~6. It is important to remark that in the case of B213-L1495 the N$_2$H$^+$ is always associated to a single C$^{18}$O component and that only single component spectra are found in this molecule.
When necessary, the integrated emission of the N$_2$H$^+$ line is also used to correct the corresponding C$^{18}$O derived column densities for depletion effects \citep[see][for a detailed discussion]{HAC13}. The distribution of the total column densities measured in all the pixels where their kinematic information is available is presented in Fig.~\ref{fig:superpos_histo} (Upper panel). As seen in these histograms, the positions surveyed in this cloud sample regions with column densities between N$(\mathrm{H}_2)\sim 1-33\times 10^{21}$~cm$^{-2}$ at a resolution of 60 arcsec. The contribution of the positions with a dense gas component detected in N$_2$H$^+$ is $\sim$~5\% of the total number of beams studied here. Errors within a factor of two in the final column densities are expected in these calculations. 

The histograms presented in  Fig.~\ref{fig:superpos_histo} (Upper panel) resemble the Column Density Probability Distribution Functions (N-PDFs) commonly employed as descriptors of the cloud structure \citep[][among others]{KAI09,LOM15,SCH15}. However, the strong observational biases (e.g., sensitivity, column density coverage, incompleteness) affecting our local C$^{18}$O observations in comparison to previous extinction or continuum based works hamper any direct comparison to these global studies (e.g., power-law behavior or log-normality of the distribution). Despite their limitations, our decomposition of the gas kinematics allows us to explore the column density distribution as a function of the number of peaks (aka superpositions) within the B213-L1495 cloud. The corresponding unfolded pseudo-N-PDFs are presented in Fig.~\ref{fig:superpos_histo} (Lower subpanels). Not surprisingly, those spectra with a higher number of peaks present statistically higher column densities than those with less multiplicity (see histograms with N$_{peaks}=1-5$). Assuming that the mean column density of the single spectra (N$_{peaks}=1$) represents the characteristic $\langle\mathrm{N}_{fiber}\rangle$, the comparison of the mean column densities at higher N$_{peaks}$ closely reproduces the expected linear $\mathrm{N}_{peaks}\times \langle\mathrm{N}_{fiber}\rangle$ relationship as a function of the number of superpositions. 

The densest components are always found at higher column densities. Interestingly, the opposite is not always guaranteed. 
Because of the aforementioned superposition of discrete structures, the highest column densities observed in this molecular cloud do not necessary map the densest components nor (only) the dense cores in the cloud. For instance, the fraction of beams with one dense component accounts for $\sim$~10\% of the beams with total column densities of N$(\mathrm{H}_2)= 1.3\times 10^{22}$~cm$^{-2}$ (or A$_V\sim13$~mag) and only becomes >70\% at relatively high values of N$(\mathrm{H}_2)= 2.3\times 10^{22}$~cm$^{-2}$ (or A$_V\sim24$~mag). Potentially, line-of-sight superpositions like those shown in Figs.~\ref{fig:complex_lines} and \ref{fig:complex2} could therefore critically affect the interpretation of extinction or continuum based N-PDFs at high column density regimes. 

The structures containing dense gas and, in particular, those associated to the positions of the dense cores identified within the B213-L1495 region are preferentially found along line-of-sights with a low number of components, that is, N$_{peaks}=$~1 or 2. Nevertheless, these dense components and cores still account for the highest column densities in these distributions. We speculate on two plausible scenarios explaining this behavior. First, although initially diffuse and with lower column densities, some of the more massive clumpy structures inside this cloud might have developed into more compact configurations either by their internal fragmentation or gravitational collapse. This ``survival'' scenario assumes that each component evolves in isolation producing a selection effect among substructures.
 Conversely, if multiple individual components coexist spatially (and not only along the line-of-sight) the large dimension of these objects (e.g., fibers) would likely produce interactions between nearby structures\footnote{Note that particularly for media with small filling factors, collisions are promoted for non-spherical geometries (e.g., filaments, sheets, etc...) due to their larger impact parameters.}. In this alternative ``merging'' scenario the densest regions would be created after the collision of several ($\ge$~2) of these substructures. Due to the likely short relaxation time after such supersonic collision (on the order of the crossing time $\tau_{cross}$), this merging of components would simultaneously produce a rapid increase of both volume and column densities in the resulting structure. The decreasing number of positions observed with large number of superposed components shown in Fig.~\ref{fig:superpos_histo} (Lower panel) would have a different origins depending on these two models. While in the first survival framework it would correspond to the statistically unlikely superposition of disconnected structures in large number, in the second merging-like scenario it would reflect the previous stages of large-scale gas collision. The lower probability of multi-object collisions and their short dissipation timescales (i.e., $\tau_{cross}$ for a single fiber) would explain the decreasing occurrence of spectra with more than 2 peaks. 

The data at hand and the limited observational access to the true 3D structure of clouds do not allow us to draw definitive conclusions about the two evolutionary scenarios presented above. Several independent lines of observational evidence do, however, favor the merging picture. First, the majority of spectra with high-multiplicity (N$_{peaks}\ge3$) detected in B213-L1495  are found in adjacent positions clustered within the B211 region (e.g., spectrum in Fig.~\ref{fig:complex_lines}). Along B213-L1495, the non-existent presence of embedded sources, as well as its low content of dense gas and CO-rich composition, identify B211 as the youngest subregion within this cloud \citep[see ][for a detailed discussion]{HAC13}. With a projected total width of $\sim$~0.1-0.5~pc, a pure line-of-sight chance alignment of the fibers identified in these spectra seems to be unlikely. Instead, these components appear to spatially coexist. Ultimately, such compact configuration might lead to collisions between several of these components. The energy dissipation of these fiber-to-fiber supersonic collisions (up to several km~s$^{-1}$) could explain localized shock fronts measured by \citet{PON14} in the mid-J (6-5) and (5-4) $^{12}$CO transitions towards kinematically rich and crowded regions like B1-East in Perseus. Additional observational and theoretical studies are necessary to explore this extreme. 

The lack of a clear substructure within the lines seems to suggest that the internal evolution of each of these fibers appear to be described by an unresolved microscopic turbulence. Within these individual objects, relatively low turbulent values of ${\cal M}\lesssim$~2-3 are found to be characteristic of this internal velocity field (e.g., Fig.~\ref{fig:complex_lines}). Local systematic variations within these ranges would be consistent with the transition to (subsonic) coherence identified in previous studies in regions like B5 in Perseus \citep{GOO98,PIN10}.

\subsection{Gas kinetic energy: micro- vs. macroscopic motions}\label{sec:KinEne}

The high line multiplicity of the $^{12}$CO and  $^{13}$CO (1-0) lines have been previously reported during the study of three high-mass star-forming regions by \citet{HEY96}. These authors pointed out that the relative motions among components are responsible of most of the observed linewidths in the total integrated spectra for each of these cloud. From the comparison of the individual linewidths with the velocity differences among these components, Heyer et al postulated that most of the kinetic energy inside these clouds resides on the relative velocities between components (macroscopic motions) instead of their internal motions (microscopic motions). Intuitively, the large number of resolved peaks with supersonic velocity differences found in Fig.~\ref{fig:complex_lines} and \ref{fig:complex2} reinforces this interpretation. 

   \begin{figure}[h]
   \centering
     \includegraphics[width=\columnwidth]{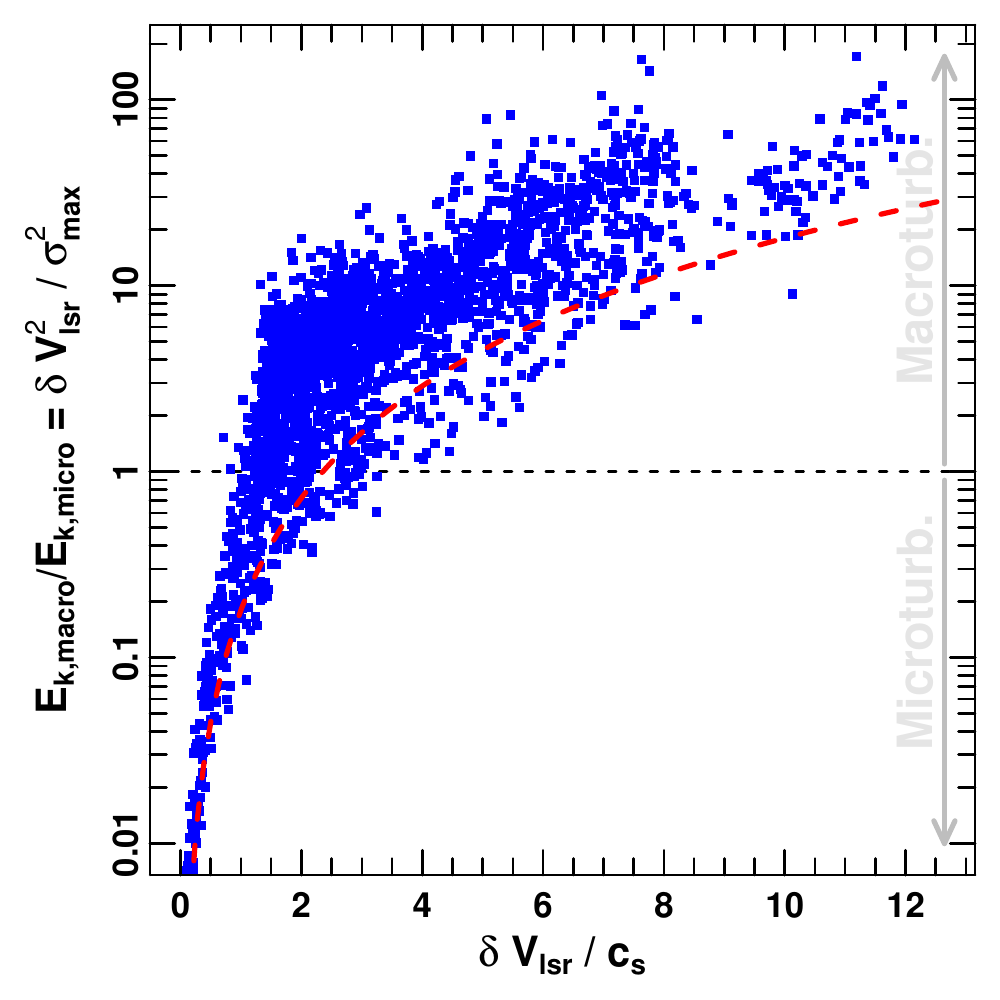}
   \caption{Ratio of the line-of-sight macro-  (E$_{kin,macro}$) and microscopic  (E$_{kin,micro}$) kinetic energies as a function of the maximum velocity differences $\delta V_{lsr}$ (in units of the sound speed at 10~K) for all the spectra observed in B213-L1495 with N$_{peaks}\ge$~2. The horizontal dashed-line at $\mathrm{E}_{kin,macro}/\mathrm{E}_{kin,micro}=1$) separates the two energy regimes dominated by microscopic ($\delta V_{lsr}^2 / \sigma_{max}^2 < 1$; $\sim$~15\% of the spectra) and macroscopic motions ($\delta V_{lsr}^2 / \sigma_{max}^2 > 1$; $\sim$~85\% of the spectra). The red dashed curve indicates the maximum velocity that can be resolved in our spectra ${\cal R}$ (see Sect.~\ref{sec:blending}).}
              \label{fig:Ek}%
    \end{figure}

We have quantified the ratio between the kinetic energies in both micro- (E$_{kin,micro}$) and macroscopic (E$_{kin,macro}$) motions along the line-of-sight by comparing the largest observed velocity dispersions $\sigma_{max}$ (without opacity corrections) with the maximum velocity differences $\delta V_{lsr}$ found in all the spectra fitted in B213-L1495 with N$_{peaks}\ge$~2 (see Sect.~\ref{sec:ColDen}). If all the measured velocity components are assumed to present equal masses, the ratio of their kinetic energies (i.e., $\mathrm{E}_{kin}=\frac{1}{2}m v^2$) can be approximated by  $\mathrm{E}_{kin,macro}/\mathrm{E}_{kin,micro} \sim \delta V_{lsr}^2 / \sigma_{max}^2$. Figure~\ref{fig:Ek} shows the estimated ratio of kinetic energy in both macro- and microscopic motions for all the spectra with multiple components in our sample. In the case of spectra with N$_{peaks}>$~2, $\delta V_{lsr}$ is referred to the maximum velocity difference found between components.
In an overwhelming 85\% of the multipeak spectra the energy budget shows $\mathrm{E}_{kin,macro}/\mathrm{E}_{kin,micro}>1$ confirming \citet{HEY96} results. With  $\langle \mathrm{E}_{kin,macro}/\mathrm{E}_{kin,micro}\rangle \simeq 9$, the average contribution to the global kinetic energy in molecular clouds from the macroscopic motions is almost an order of magnitude larger than the microscopic component. These findings qualitatively resemble previous observational results indicating that most of the kinetic energy in molecular clouds resides on large-scales \citep[e.g.][]{BRU02}.

By comparing their kinetic and gravitational energies, the virial parameter, $\alpha=5\sigma ^2 R / G M$, determines the physical state of molecular clouds \citep{BER92}: Subcritical clouds ($\alpha > 2$) are supported by their internal kinetic energy while supercritical clouds ($\alpha < 2$) inevitably collapse by their own self-gravity \cite[see][for a review]{KAU13}. Estimates of the virial parameter are carried out from the comparison of the total CO intensities (i.e., total Mass) and linewidths (or $\sigma$) in molecular clouds \citep{SCO87,SOL87,ROM10}.  
However, the broadest and highly supersonic linewidths observed in CO do not reflect an internal total random motion and velocity dispersion within the cloud. If so, and as demonstrated in previous sections for single components, the CO linewidths would be dominated by optically thick effects leading to highly overestimated virial parameters ($\sigma_{int}\ll \sigma$). Instead, they are mainly produced by the superposition of multiple structures with relative supersonic orbital motions. At large scales, the virial parameter is therefore dominated by the kinetic energy between individual (macroscopic) structures. Due to the low filling factor of these structures, the kinetic energy at large scales should mainly occur at discretized values (E$_{kin}\propto \delta V_{lsr}^2 = (V_{lsr,1}-V_{lsr,2})^2$). 

Although minor in the total energy budget, the transonic internal motions of the individual components become important at smaller scales. Acting as microturbulence, these unresolved motions provide internal support against the local gravitational collapse. 
This kinematic organization combining large-scale supersonic velocity differences and local random sonic-like motions facilitates the clouds to present different virial parameters depending on the scale. While globally subcritically stabilized by large-scale macrotubulent motions, clouds could simultaneously contain supercritical small pockets of gas if their internal velocity dispersion is not able to counteract the local gravity.

\section{Conclusions}

We have studied the molecular emission of the three main CO isotopologues, namely  $^{12}$CO, $^{13}$CO, and C$^{18}$O, in different molecular clouds. By combining both observations and simplified LTE simulations we have investigated their emission line profiles and, in particular, the opacity effects in the interpretation of these lines. The main results of this study are the following:

\begin{enumerate}
\item The observed FWHM of the fundamental J=1-0 transition of highly abundant $^{12}$CO and $^{13}$CO tracers are systematically affected by opacity broadening effects. The observed FWHM of these high opacity lines reflect the intensity increase of their less opaque line wings rather than the true gas velocity field along the line-of-sight. More than 90\% of the observed differences between these tracers are consistent with a pure opacity broadening. 

\item As a result of this opacity effects, the velocity dispersions directly derived from the observed $^{12}$CO and $^{13}$CO linewidths typically overestimate the intrinsic velocity dispersion of the gas up to a factor 2-3 for individual velocity components. Opacity corrections should be then applied in the study of these lines.

\item Line blending effects produced by the combination of high optical depths with multiple narrow velocity-components appears as an effective mechanism creating broad, several km~s$^{-1}$-width, $^{12}$CO and $^{13}$CO lines. The analysis of these line profiles requires to take into account the line velocity structure observed in less abundant isotopologues like C$^{18}$O.

\item While usually assumed as intrinsically single supersonic lines, these broad profiles might reveal a superposition of multiple (tran-)sonic line velocity components separated by supersonic velocity differences. In contrast to the classical interpretation in the framework of the unresolved, microscopic turbulence, these findings would suggest that the internal dynamics of molecular clouds would be described by a macroscopic turbulence as for example generated in molecular clouds that form in a cooling converging gas flows \citep[e.g.][]{HEI08a}. 
These numerical simulations indeed lead to a highly clumpy cold component with intrinsically subsonic turbulence, embedded in a more diffuse and warm inter-clump medium. The clumps move supersonically through the cloud environment, when considering their intrinsic sound speed. Their motion is however trans-sonic with respect to the sound speed of the inter-clump medium.

\item These emission characteristics of the CO lines reported in this work suggests that clouds might be internally created by a complex organization of fibers. Independently of their shape, these structures would be defined as continuous and coherent structures both spatially and in velocity. Tangled and nested inside clouds, these fibers would dominate the bulk of the molecular mass detected in CO. Most of the gas kinematics and internal supersonic motions inside clouds would be then created by the macroscopic and discretized motions between these objects.

\end{enumerate}

\begin{acknowledgements}
      The authors thank Mark Heyer for kindly sharing FCRAO data.
      A.H. gratefully acknowledges support from Ewine van Dishoeck, John Tobin, and Magnus Persson during his stay at the Leiden Observatory. 
      A.H. thanks the insightful discussions and comments from Mario Tafalla and Jan Forbrich.
      This publication is supported by the Austrian Science Fund (FWF).
      This work was supported by the cluster of excellence ``Origin and Structure of the Universe''.
      This work was carried out in part at the Jet Propulsion Laboratory, which is operated by the California Institute of Technology for NASA.
 \end{acknowledgements}


\begin{thebibliography}{}
\bibitem[Andr{\'e} et al.(2010)]{AND10} Andr{\'e}, P., Men'shchikov, A., Bontemps, S., et al.\ 2010, \aap, 518, L102 
\bibitem[Arzoumanian et al.(2013)]{ARZ13} Arzoumanian, D., Andr{\'e}, P., Peretto, N., K\"onyves, V.\ 2013, \aap, 553, A119 
\bibitem[Ballesteros-Paredes et al.(1999)]{BAL99} Ballesteros-Paredes, J., V{\'a}zquez-Semadeni, E., \& Scalo, J.\ 1999, \apj, 515, 286 
\bibitem[Bally \& Langer(1982)]{BAL82} Bally, J., \& Langer, W.~D.\ 1982, \apj, 255, 143 
\bibitem[Baker(1976)]{BAK76} Baker, P.~L.\ 1976, \aap, 50, 327 
\bibitem[Beaumont et al.(2013)]{BEA13} Beaumont, C.~N., Offner, S.~S.~R., Shetty, R., Glover, S.~C.~O., \& Goodman, A.~A.\ 2013, \apj, 777, 173 
\bibitem[Bertoldi \& McKee(1992)]{BER92} Bertoldi, F., \& McKee, C.~F.\ 1992, \apj, 395, 140 
\bibitem[Bolatto et al.(2008)]{BOL08} Bolatto, A.~D., Leroy, A.~K., Rosolowsky, E., Walter, F., \& Blitz, L.\ 2008, \apj, 686, 948 
\bibitem[Buckle et al.(2012)]{BUC12} Buckle, J.~V., Davis, C.~J., Francesco, J.~D., et al.\ 2012, \mnras, 422, 521 
\bibitem[Burkert \& Hartmann(2004)]{BUR04} Burkert, A., \& Hartmann, L.\ 2004, \apj, 616, 288 
\bibitem[Brunt \& Heyer(2002)]{BRU02} Brunt, C.~M., \& Heyer, M.~H.\ 2002, \apj, 566, 289 
\bibitem[Chu \& Watson(1983)]{CHU83} Chu, Y.-H., \& Watson, W.~D.\ 1983, \apj, 267, 151 
\bibitem[Correia et al.(2014)]{COR14} Correia, C., Burkhart, B., Lazarian, A., et al.\ 2014, \apjl, 785, L1 
\bibitem[Dame et al.(2001)]{DAM01} Dame, T.~M., Hartmann, D., \& Thaddeus, P.\ 2001, \apj, 547, 792
\bibitem[Dickman \& Clemens(1983)]{DIC83} Dickman, R.~L., \& Clemens, D.~P.\ 1983, \apj, 271, 143 
\bibitem[Dobbs et al.(2012)]{DOB12} Dobbs, C.~L., Pringle, J.~E., \& Burkert, A.\ 2012, \mnras, 425, 2157
\bibitem[Dobbs et al.(2014)]{DOB14} Dobbs, C.~L., Krumholz, M.~R., Ballesteros-Paredes, J., et al.\ 2014, Protostars and Planets VI, 3
\bibitem[Elmegreen \& Scalo(2004)]{ELM04} Elmegreen, B.~G., \& Scalo, J.\ 2004, \araa, 42, 211 
\bibitem[Evans(1999)]{EVA99} Evans, N.~J., II 1999, \araa, 37, 311
\bibitem[Falgarone \& Phillips(1990)]{FAL90} Falgarone, E., \& Phillips, T.~G.\ 1990, \apj, 359, 344 
\bibitem[Falgarone et al.(1991)]{FAL91} Falgarone, E., Phillips, T.~G., \& Walker, C.~K.\ 1991, \apj, 378, 186  
\bibitem[Frerking et al.(1982)]{FRE82} Frerking, M.~A., Langer, W.~D., \& Wilson, R.~W.\ 1982, \apj, 262, 590 
\bibitem[Goldsmith \& Langer(1999)]{GOL99} Goldsmith, P.~F., \& Langer, W.~D.\ 1999, \apj, 517, 209 
\bibitem[Goldsmith et al.(2008)]{GOL08} Goldsmith, P.~F., Heyer, M., Narayanan, G., et al.\ 2008, \apj, 680, 428 
\bibitem[Goldsmith(2013)]{GOL13} Goldsmith, P.~F.\ 2013, \apj, 774, 134 
\bibitem[Goodman et al.(1998)]{GOO98} Goodman, A.~A., Barranco, J.~A., Wilner, D.~J., \& Heyer, M.~H.\ 1998, \apj, 504, 223 
\bibitem[Grenier et al.(2005)]{GRE05} Grenier, I.~A., Casandjian, J.-M., \& Terrier, R.\ 2005, Science, 307, 1292 
\bibitem[Greve et al.(2005)]{GRA05} Greve, T.~R., Bertoldi, F., Smail, I., et al.\ 2005, \mnras, 359, 1165 
\bibitem[Hacar \& Tafalla(2011)]{HAC11} Hacar, A., \& Tafalla, M.\ 2011, \aap, 533, A34 
\bibitem[Hacar et al.(2013)]{HAC13} Hacar, A., Tafalla, M., Kauffmann, J., \& Kov{\'a}cs, A.\ 2013, \aap, 554, A55
\bibitem[Hacar et al.(2015)]{HAC15} Hacar, A., Kainulainen, J., Tafalla, M., Beuther, H., \& Alves, J.\ 2015, arXiv:1511.06370 
\bibitem[Hartmann \& Burkert(2007)]{HAR07} Hartmann, L., \& Burkert, A.\ 2007, \apj, 654, 988 
\bibitem[Heyer et al.(1987)]{HEY87} Heyer, M.~H., Vrba, F.~J., Snell, R.~L., et al.\ 1987, \apj, 321, 855 
\bibitem[Heyer et al.(1996)]{HEY96} Heyer, M.~H., Carpenter, J.~M., \& Ladd, E.~F.\ 1996, \apj, 463, 630
\bibitem[Heitsch et al.(2008)]{HEI08a} Heitsch, F., Hartmann, L.~W., \& Burkert, A.\ 2008, \apj, 683, 786
\bibitem[Heitsch et al.(2008)]{HEI08b} Heitsch, F., Hartmann, L.~W., Slyz, A.~D., Devriendt, J.~E.~G., \& Burkert, A.\ 2008, \apj, 674, 316 
\bibitem[Kainulainen et al.(2009)]{KAI09} Kainulainen, J., Beuther, H., Henning, T., \& Plume, R.\ 2009, \aap, 508, L35 
\bibitem[Kauffmann et al.(2013)]{KAU13} Kauffmann, J., Pillai, T., \& Goldsmith, P.~F.\ 2013, \apj, 779, 185 
\bibitem[Keto \& Myers(1986)]{KET86} Keto, E.~R., \& Myers, P.~C.\ 1986, \apj, 304, 466 
\bibitem[Klessen et al.(2000)]{KLE00} Klessen, R.~S., Heitsch, F., \& Mac Low, M.-M.\ 2000, \apj, 535, 887 
\bibitem[Kwan \& Scoville(1976)]{KWA76} Kwan, J., \& Scoville, N.\ 1976, \apjl, 210, L39 
\bibitem[Kwan \& Sanders(1986)]{KWA86} Kwan, J., \& Sanders, D.~B.\ 1986, \apj, 309, 783 
\bibitem[Ladd \& Myers(1991)]{LAD91} Ladd, E.~F., \& Myers, P.~C.\ 1991, Atoms, Ions and Molecules: New Results in Spectral Line Astrophysics, 16, 241 
\bibitem[Langer et al.(1984)]{LAN84} Langer, W.~D., Graedel, T.~E., Frerking, M.~A., \& Armentrout, P.~B.\ 1984, \apj, 277, 581
\bibitem[Larson(1981)]{LAR81} Larson, R.~B.\ 1981, \mnras, 194, 809
\bibitem[Leroy et al.(2008)]{LER08} Leroy, A.~K., Walter, F., Brinks, E., et al.\ 2008, \aj, 136, 2782 
\bibitem[Leung(1978)]{LEN78} Leung, C.~M.\ 1978, \apj, 225, 427
\bibitem[Li \& Goldsmith(2003)]{LI03} Li, D., \& Goldsmith, P.~F.\ 2003, \apj, 585, 823 
\bibitem[Liszt et al.(1974)]{LIS74} Liszt, H.~S., Wilson, R.~W., Penzias, A.~A., et al.\ 1974, \apj, 190, 557
\bibitem[Lombardi et al.(2010)]{LOM10} Lombardi, M., Lada, C.~J., \& Alves, J.\ 2010, \aap, 512, A67 
\bibitem[Lombardi et al.(2015)]{LOM15} Lombardi, M., Alves, J., \& Lada, C.~J.\ 2015, \aap, 576, L1 
\bibitem[Lynds(1962)]{LYN62} Lynds, B.~T.\ 1962, \apjs, 7, 1 
\bibitem[Mac Low \& Ossenkopf(2000)]{MAC00} Mac Low, M.-M., \& Ossenkopf, V.\ 2000, \aap, 353, 339
\bibitem[Mac Low \& Klessen(2004)]{MAC04} Mac Low, M.-M., \& Klessen, R.~S.\ 2004, Reviews of Modern Physics, 76, 125 
\bibitem[Miesch \& Bally(1994)]{MIE94} Miesch, M.~S., \& Bally, J.\ 1994, \apj, 429, 645
\bibitem[Moeckel \& Burkert(2014)]{MOC14} Moeckel, N., \& Burkert, A.\ 2014, arXiv:1402.2614 
\bibitem[Myers et al.(1983)]{MYE83} Myers, P.~C., Linke, R.~A., \& Benson, P.~J.\ 1983, \apj, 264, 517
\bibitem[Narayanan et al.(2008)]{NAR08} Narayanan, G., Heyer, M.~H., Brunt, C., et al.\ 2008, \apjs, 177, 341 
\bibitem[Ntormousi et al.(2011)]{NTO11} Ntormousi, E., Burkert, A., Fierlinger, K., \& Heitsch, F.\ 2011, \apj, 731, 13
\bibitem[Onishi et al.(1996)]{ONI96} Onishi, T., Mizuno, A., Kawamura, A., Ogawa, H., \& Fukui, Y.\ 1996, \apj, 465, 815 
\bibitem[Ossenkopf et al.(2001)]{OSS01} Ossenkopf, V., Klessen, R.~S., \& Heitsch, F.\ 2001, \aap, 379, 1005
\bibitem[Ossenkopf \& Mac Low(2002)]{OSS02} Ossenkopf, V., \& Mac Low, M.-M.\ 2002, \aap, 390, 307
\bibitem[Palmeirim et al.(2013)]{PAL13} Palmeirim, P., Andr{\'e}, P., Kirk, J., et al.\ 2013, \aap, 550, A38 
\bibitem[Panopoulou et al.(2014)]{PAN14} Panopoulou, G.~V., Tassis, K., Goldsmith, P.~F., \& Heyer, M.~H.\ 2014, \mnras, 444, 2507 
\bibitem[Park et al.(2004)]{PAR04} Park, Y.-S., Lee, C.~W., \& Myers, P.~C.\ 2004, \apjs, 152, 81 
\bibitem[Penzias et al.(1971)]{PEN71} Penzias, A.~A., Jefferts, K.~B., \& Wilson, R.~W.\ 1971, \apj, 165, 229
\bibitem[Phillips et al.(1973)]{PHI73} Phillips, T.~G., Jefferts, K.~B., \& Wannier, P.~G.\ 1973, \apjl, 186, L19 
\bibitem[Phillips et al.(1979)]{PHI79} Phillips, T.~G., Huggins, P.~J., Wannier, P.~G., \& Scoville, N.~Z.\ 1979, \apj, 231, 720 
\bibitem[Pineda et al.(2008)]{PIN08} Pineda, J.~E., Caselli, P., \& Goodman, A.~A.\ 2008, \apj, 679, 481 
\bibitem[Pineda et al.(2010)]{PIN10} Pineda, J.~L., Goldsmith, P.~F., Chapman, N., et al.\ 2010, \apj, 721, 686 
\bibitem[Pineda et al.(2013)]{PIN13} Pineda, J.~L., Langer, W.~D., Velusamy, T., \& Goldsmith, P.~F.\ 2013, \aap, 554, A103
\bibitem[Pon et al.(2014)]{PON14} Pon, A., Johnstone, D., Kaufman, M.~J., Caselli, P., \& Plume, R.\ 2014, \mnras, 445, 1508 
\bibitem[Rickard et al.(1975)]{RIC75} Rickard, L.~J., Palmer, P., Morris, M., Zuckerman, B., \& Turner, B.~E.\ 1975, \apjl, 199, L75 
\bibitem[Robert \& Pagani(1993)]{ROB93} Robert, C., \& Pagani, L.\ 1993, \aap, 271, 282 
\bibitem[Rohlfs \& Wilson(1996)]{ROL96} Rohlfs, K., \& Wilson, T.~L.\ 1996, Tools of Radio Astronomy, XVI, 423 pp.~127 figs., 20 tabs..~ Springer-Verlag Berlin Heidelberg New York.~Also Astronomy and Astrophysics Library,  
\bibitem[Roman-Duval et al.(2010)]{ROM10} Roman-Duval, J., Jackson, J.~M., Heyer, M., Rathborne, J., \& Simon, R.\ 2010, \apj, 723, 492 
\bibitem[Sadavoy et al.(2015)]{SAD15} Sadavoy, S.~I., Shirley, Y., Di Francesco, J., et al.\ 2015, arXiv:1504.05206 
\bibitem[Sage et al.(1991)]{SAG91} Sage, L.~J., Henkel, C., \& Mauersberger, R.\ 1991, \aap, 249, 31 
\bibitem[Schmalzl et al.(2010)]{SCH10} Schmalzl, M., Kainulainen, J., Quanz, S.~P., et al.\ 2010, \apj, 725, 1327
\bibitem[Schneider et al.(2015)]{SCH15} Schneider, N., Bontemps, S., Motte, F., et al.\ 2015, arXiv:1509.01082
\bibitem[Scoville \& Solomon(1974)]{SCO74} Scoville, N.~Z., \& Solomon, P.~M.\ 1974, \apjl, 187, L67 
\bibitem[Scoville et al.(1987)]{SCO87} Scoville, N.~Z., Yun, M.~S., Sanders, D.~B., Clemens, D.~P., \& Waller, W.~H.\ 1987, \apjs, 63, 821 
\bibitem[Shirley(2015)]{SHI15} Shirley, Y.~L.\ 2015, \pasp, 127, 299
\bibitem[Smith et al.(2014)]{SMI14} Smith, R.~J., Glover, S.~C.~O., \& Klessen, R.~S.\ 2014, \mnras, 445, 2900 
\bibitem[Smith et al.(2016)]{SMI16} Smith, R.~J., Glover, S.~C.~O., Klessen, R.~S., \& Fuller, G.~A.\ 2016, \mnras, 455, 3640
\bibitem[Snell et al.(1980)]{SNE80} Snell, R.~L., Loren, R.~B., \& Plambeck, R.~L.\ 1980, \apjl, 239, L17 
\bibitem[Solomon \& de Zafra(1975)]{SOL75} Solomon, P.~M., \& de Zafra, R.\ 1975, \apjl, 199, L79 
\bibitem[Solomon et al.(1987)]{SOL87} Solomon, P.~M., Rivolo, A.~R., Barrett, J., \& Yahil, A.\ 1987, \apj, 319, 730
\bibitem[Solomon et al.(1997)]{SOL97} Solomon, P.~M., Downes, D., Radford, S.~J.~E., \& Barrett, J.~W.\ 1997, \apj, 478, 144 
\bibitem[Tafalla et al.(1998)]{TAF98} Tafalla, M., Mardones, D., Myers, P.~C., et al.\ 1998, \apj, 504, 900 
\bibitem[Tafalla \& Hacar(2015)]{TAF15} Tafalla, M., \& Hacar, A.\ 2015, \aap, 574, A104 
\bibitem[Tauber et al.(1991)]{TAU91} Tauber, J.~A., Goldsmith, P.~F., \& Dickman, R.~L.\ 1991, \apj, 375, 635 
\bibitem[Ungerechts \& Thaddeus(1987)]{UNG87} Ungerechts, H., \& Thaddeus, P.\ 1987, \apjs, 63, 645 
\bibitem[van Dishoeck \& Black(1988)]{VDH88} van Dishoeck, E.~F., \& Black, J.~H.\ 1988, \apj, 334, 771 
\bibitem[van Dishoeck et al.(1991)]{VDH91} van Dishoeck, E.~F., Black, J.~H., Phillips, T.~G., \& Gredel, R.\ 1991, \apj, 366, 141 
\bibitem[V{\'a}zquez-Semadeni et al.(2007)]{VAZ07} V{\'a}zquez-Semadeni, E., G{\'o}mez, G.~C., Jappsen, A.~K., et al.\ 2007, \apj, 657, 870
\bibitem[Vilas-Boas et al.(1994)]{VIL94} Vilas-Boas, J.~W.~S., Myers, P.~C., \& Fuller, G.~A.\ 1994, \apj, 433, 96 
\bibitem[Vilas-Boas et al.(2000)]{VIL00} Vilas-Boas, J.~W.~S., Myers, P.~C., \& Fuller, G.~A.\ 2000, \apj, 532, 1038 
\bibitem[White et al.(2015)]{WHI15} White, G.~J., Drabek-Maunder, E., Rosolowsky, E., et al.\ 2015, \mnras, 447, 1996
\bibitem[Wilson et al.(1970)]{WIL70} Wilson, R.~W., Jefferts, K.~B., \& Penzias, A.~A.\ 1970, \apjl, 161, L43 
\bibitem[Wilson \& Rood(1994)]{WIL94} Wilson, T.~L., \& Rood, R.\ 1994, \araa, 32, 191
\bibitem[Wilson et al.(1999)]{WIL99} Wilson, C.~D., Howe, J.~E., \& Balogh, M.~L.\ 1999, \apj, 517, 174
\bibitem[Wolfire et al.(1993)]{WOL93} Wolfire, M.~G., Hollenbach, D., \& Tielens, A.~G.~G.~M.\ 1993, \apj, 402, 195 
\bibitem[Wolfire et al.(2010)]{WOL10} Wolfire, M.~G., Hollenbach, D., \& McKee, C.~F.\ 2010, \apj, 716, 1191 
\bibitem[Wong et al.(2008)]{WON08} Wong, T., Ladd, E.~F., Brisbin, D., et al.\ 2008, \mnras, 386, 1069
\bibitem[Young et al.(1982)]{YOU82} Young, J.~S., Goldsmith, P.~F., Langer, W.~D., Wilson, R.~W., \& Carlson, E.~R.\ 1982, \apj, 261, 513 
\bibitem[Zuckerman \& Evans(1974)]{ZUC74} Zuckerman, B., \& Evans, N.~J., II 1974, \apjl, 192, L149 
\end{thebibliography}
\end{document}